\documentclass[11pt,draftcls,onecolumn]{IEEEtran}
\usepackage{}

\usepackage{amsfonts}
\usepackage{amssymb}
\usepackage{mathrsfs}
\usepackage{amsmath}
\usepackage{cite}
\usepackage{mathrsfs}
\usepackage[ruled,vlined]{algorithm2e}
\usepackage{pgf}
\usepackage{tikz}
\usetikzlibrary{arrows,automata}
\usepackage[latin1]{inputenc}
\usepackage{verbatim}
\makeatletter

\newcommand{\Rmnum}[1]{\expandafter\@slowromancap\romannumeral #1@}
\makeatother

\newtheorem{thm}{Theorem}
\newtheorem{lemma}[thm]{Lemma}
\newtheorem{eg}{Example}
\newtheorem{prop}{Proposition}

\newtheorem{defn}{Definition}
\newtheorem{remark}{Remark}

\newtheorem{rem-eg}[thm]{Remark and Example}

\newcommand{\rom}{\romannumeral}

\SetKw{KwInitialization}{Initialization:}

\begin{document}
%
\title{Distributed Storage over Unidirectional Ring Networks
\thanks{
This research is supported by the National Key Basic Research Program of China (973 Program Grant No. 2013CB834204), the National Natural Science Foundation of China (Nos. 61301137, 61171082) and the Fundamental Research Funds for Central Universities of China (No. 65121007).
}}

\author{Jiyong~Lu,
        Xuan~Guang and
        Fang-Wei~Fu
\thanks{J. Lu is with the Chern Institute of Mathematics, Nankai University, Tianjin 300071, P. R. China. Email: lujiyong@mail.nankai.edu.cn.}
\thanks{X. Guang is with the School of Mathematical Science and LPMC, Nankai University, Tianjin 300071, P. R. China. Email: xguang@nankai.edu.cn.}
\thanks{F.-W. Fu is with the Chern Institute of Mathematics and LPMC, Nankai University, Tianjin 300071, P. R. China. Email: fwfu@nankai.edu.cn.}}

\markboth{Distributed Storage over Unidirectional Ring Networks}%
{Guang \MakeLowercase{\textit{et al.}}: Distributed Storage over Unidirectional Ring Networks}

\maketitle

\begin{abstract}
In this paper, we study distributed storage problems over unidirectional ring networks, whose storage nodes form a directed ring and data is transmitted along the same direction. The original data is distributed to store on these nodes. Each user can connect one and only one storage node to download the total original data. A lower bound on the reconstructing bandwidth to recover the original data for each user is proposed, and it is achievable for arbitrary parameters. If a distributed storage scheme can achieve this lower bound with equality for every user, we say it an optimal reconstructing distributed storage scheme (ORDSS). Furthermore, the repair problem for a failed storage node in ORDSSes is under consideration and a tight lower bound on the repair bandwidth is obtained. In particular, we indicate the fact that for any ORDSS, every failed storage node can be repaired with repair bandwidth satisfying the lower bound with equality. In addition, we present two constructions for ORDSSes of arbitrary parameters, called MDS construction and ED construction, respectively. Particularly, ED construction, using the concept of Euclidean division, is more efficient by our analysis in detail.
\end{abstract}


%
\IEEEpeerreviewmaketitle

\section{Introduction}

\IEEEPARstart{D}{istributed} storage systems can keep data reliable over unreliable storage nodes for a long period. To ensure the reliability, redundancy has to be introduced. All kinds of strategies have been proposed to generate redundancy, such as replication \cite{Bolosky-etc-2000}, \cite{Rowstron-etc-2001}, erasure codes \cite{M-J3-1995}-\cite{Lin-Tzeng-2010}, and regenerating codes \cite{Dimakis-etc-2010}, \cite{Wu-Dimakis-2009}, etc.
Both erasure codes and regenerating codes keep MDS property for data reconstruction, that is, arbitrary $k$ out of $n$ storage nodes can reconstruct the original data. Moreover, for a regenerating code, when a storage node fails, we can repair it by connecting to $d$ $(\geq k)$ remaining storage nodes and downloading $\beta$ symbols from each one. This process is called repair process and the total amount $d\beta$ of data downloaded for repairing this failed node is termed as repair bandwidth. In general, the repair bandwidth $d\beta$ is less than the size $M$ of the original data. Motivated by network coding \cite{L-Y-C}, Dimakis \textit{et al.} \cite{Dimakis-etc-2010} used information flow graphs to express regenerating codes and showed that regenerating codes could be designed by random linear network codes \cite{T-ect}. Further, by the cut-set bound of network coding, they established a relation among the parameters of regenerating codes as follows:
$$M \leq \sum^{k-1}_{i=0}{\rm min} \{(d-i)\beta,\alpha\}.$$ From the above inequality, we can deduce that $\alpha$ and $\beta$ are not possible to be minimized simultaneously and thus there is a tradeoff between choices of $\alpha$ and $\beta$. In particular, the two extreme points in this tradeoff correspond to the minimum storage regenerating (MSR) codes and the minimum bandwidth regenerating (MBR) codes, respectively, which have been studied widely. For example, the constructions of MBR codes for all parameters $(n,k,d)$, $k\leq d\leq n-1$, and MSR codes for all $2k-2\leq d\leq n-1$ were presented in \cite{Rashmi-Shah-Kumar-2011}. Lin and Chung \cite{Lin-Chung-2014} recently discussed the novel repair-by-transfer codes for MBR points, in which the remaining storage nodes for repairing a failed node only need to pass a portion of the stored symbols without any arithmetic operations. Cooperative regenerating codes were presented firstly by Hu \textit{et al.} \cite{Hu-etc-2010} to repair multiple failed storage nodes simultaneously, where information exchanges among new substituted nodes are allowed. Cooperative regenerating codes can be regarded as an extension of regenerating codes which can just repair failed storage nodes one by one. The authors also obtained a lower bound on the repair bandwidth. For more discussions about this topic, please refer to \cite{Shum-2011}, \cite{Scouarnec-2012}.

A large number of constructions for distributed storage schemes do not concern network structures among storage nodes. Actually, in many applications, storage nodes have certain topological relationships, such as the hierarchical network structure \cite{Benson-Akella-Maltz-2010}, the multi-hop network structure \cite{Kong-Aly-Soljanin-2010}, and so on. It is not difficult to see that network structures possibly make effect to reconstructing and repair processes. Li \textit{et al.} \cite{Li-etc-2010} studied repair-time in tree-structure networks which have links with different capacities. In \cite{Shum-etc}, Gerami \textit{et al.} considered repair-cost in multi-hop networks and formulated the minimum-cost as a linear programming problem for linear costs. Inspired by these works, in this paper, we focus on distributed storage problems over a class of simple but important networks, unidirectional ring networks, which usually exist as a part of complex networks. In these unidirectional ring networks, storage nodes form a directed ring and data is transmitted along the same direction. Each user can connect one and only one storage node to download data. For each user, its reconstructing bandwidth  to recover the original data is the total number of transmitted symbols. Particularly, when the cost of transmitting a symbol on different edges is regarded to be the same and denoted as 1, then our reconstructing bandwidth actually is equal to reconstructing cost. By cut-set bound analysis of the corresponding information flow graph, we obtain a lower bound on the reconstructing bandwidth and further indicate its tightness for arbitrary parameters. For a distributed storage scheme, we say it an optimal reconstructing distributed storage scheme (ORDSS), if the reconstructing bandwidth for every user satisfies this lower bound with equality. Furthermore, we study the repair problem of ORDSSes, and also deduce a tight lower bound on the repair bandwidth, which is the total number of transmitted symbols to repair a failed storage node. Particularly, we show that every ORDSS can satisfy this lower bound with equality. In addition, we present two constructions of ORDSSes, called MDS construction and ED construction, respectively. The first construction uses MDS codes in the algebraic coding theory and the ED construction applies the concept of Euclidean division. Both of them can be used for arbitrary parameters. However, the MDS construction has some shortcomings. For example, when the parameters take large values, the finite field size and computational complexity for the MDS construction will be too large for practical applications. This is because the MDS property is too strong for constructions of ORDSSes, although it can solve our proposed constructing problem. While the ED construction always uses the smallest finite field $\mathbb{F}_2$, so it can make up those defects of MDS construction and shows good performance.

The remainder of the paper is organized as follows. In Section \ref{ME-BM}, we first propose a motivating example to illustrate our research problems, and then describe the basic mathematical model of these problems over unidirectional ring networks. Tight lower bounds on the reconstructing and repair bandwidths are discussed in detail in Section \ref{B-RRB}. Moreover, the MDS construction approach of ORDSSes is also implied. In Section \ref{cons}, by the concept of Euclidean division, we present another construction approach of ORDSSes for arbitrary parameters $(n,\alpha,M)$, called ED construction. Particularly, we compare the two constructions and show that the ED construction is more efficient than the MDS construction. Finally, the paper is concluded in Section \ref{conc}.

\section{Preliminaries}\label{ME-BM}
In this section, we first give an example to show our research problems. This also implies that it is meaningful and interesting to study distributed storage problems over unidirectional ring networks. Furthermore, we propose the basic mathematical model of our research problems and introduce some notation and definitions.
\subsection{A Motivating Example}\label{ME}
\begin{figure}[!ht]
\centering
\includegraphics[width=6.5cm,height=3.9cm]{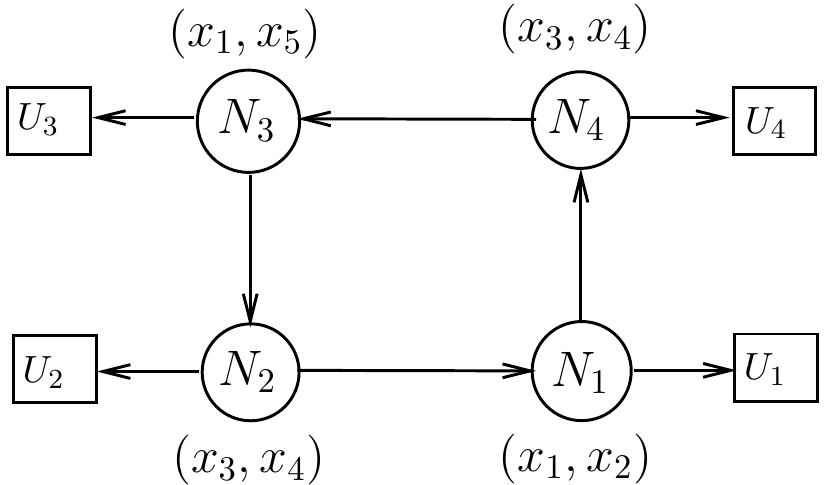}
\caption{The unidirectional ring network with $n=4$, $\alpha=2$, $M=5$.}
\label{fig_cn}
\end{figure}

Fig. \ref{fig_cn} depicts a unidirectional ring network with $n=4$ storage nodes, denoted by $N_1,N_2,N_3,N_4$. The data exchanges between the storage nodes
along the given direction of the ring network.
Each storage node has storage capacity $\alpha=2$. Let the row vector of original data be $X=[x_1,x_2,x_3,x_4,x_5]\in\mathbb{F}^5_5$, that is, the size $M$ of the original data is 5. All four storage nodes distributed store the original data $X$. Each user can connect one and only one storage node to download the original data. Without loss of generality, let user node $U_i$ connect storage node $N_i$, $1\leq i\leq 4$. Fig. \ref{fig_cn} also gives a distributed storage scheme, in which every user can reconstruct the original data $X$. For instances, $N_1$ stores $x_1$ and $x_2$, $N_2$ stores $x_3$ and $x_4$.
\begin{figure}[!ht]
\centering
\includegraphics[width=11cm,height=4.3cm]{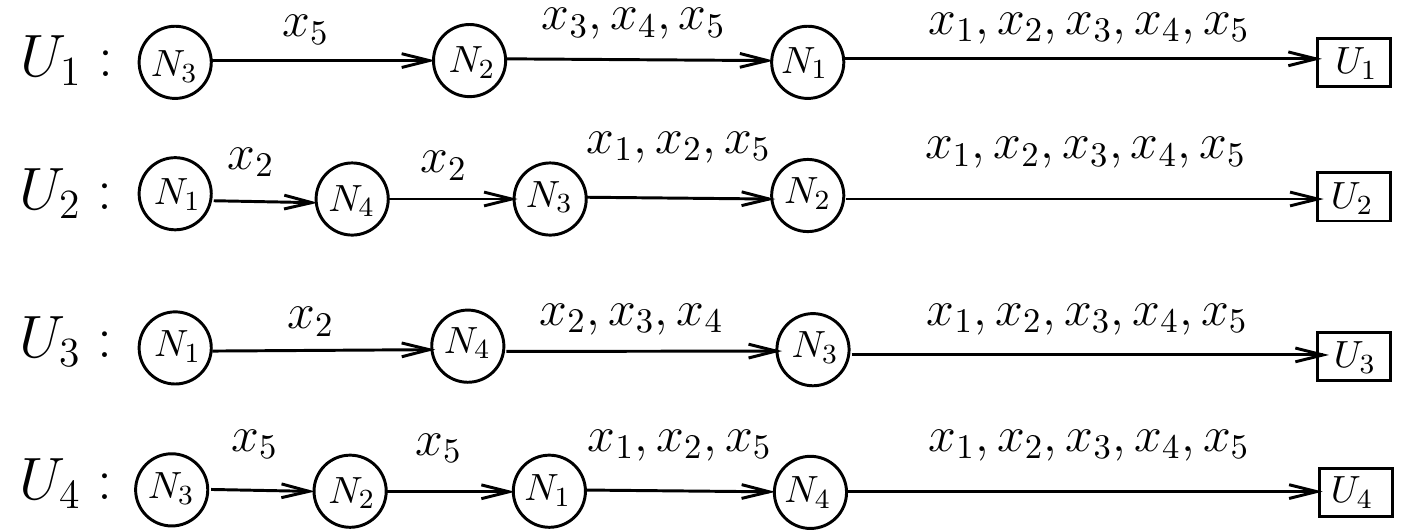}
\caption{The reconstructing process for all users. Each of $U_1,U_3$ obtains the original data with reconstructing bandwidth 9, while each of $U_2,U_4$ with reconstructing bandwidth 10.}
\label{fig_rc}
\end{figure}
For this scheme, Fig. \ref{fig_rc} describes an optimal reconstructing process which minimizes the reconstructing bandwidth for every user.
For example, in order to reconstruct the original data at the user node $U_1$, $N_3$ transmits $x_5$ to the storage node $N_2$, $N_2$ transmits three symbols $x_3,x_4,x_5$ to the storage node $N_1$, then together with its own stored symbols $x_1,x_2$, $N_1$ transmits all the original symbols $x_1,x_2,x_3,x_4,x_5$ to the user $U_1$. Thus, $U_1$ can obtain the original data $X$. Based on this reconstructing process, we know that the minimum reconstructing bandwidth on average of this scheme is $38/4=9.5$. Naturally, we propose a series of problems as follows: does there exist a distributed storage scheme with the minor reconstructing bandwidth? what is the minimum of the reconstructing bandwidth? how to efficiently construct distributed storage schemes achieving the minimum reconstructing bandwidth?

\begin{figure}[!ht]
\centering
\includegraphics[width=6.5cm,height=3.9cm]{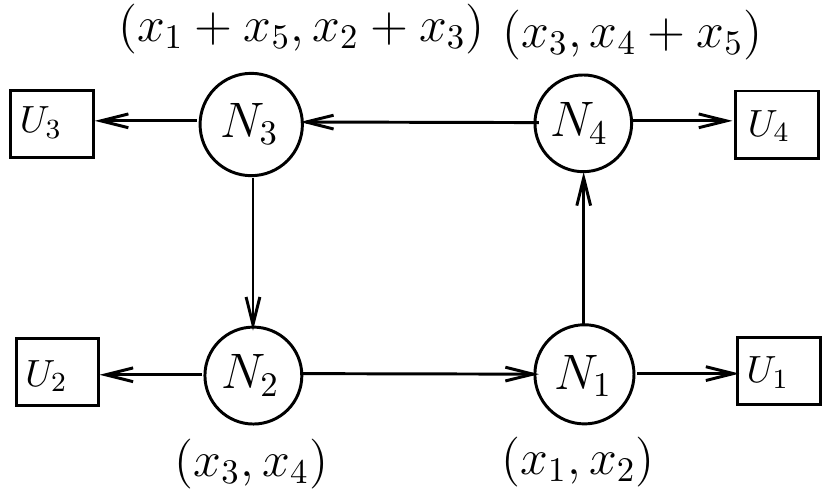}
\caption{The new storage scheme with $n=4,\alpha=2,M=5$.}
\label{fig_mrc}
\end{figure}

\begin{figure}[!ht]
\centering
\includegraphics[width=10.5cm,height=4.5cm]{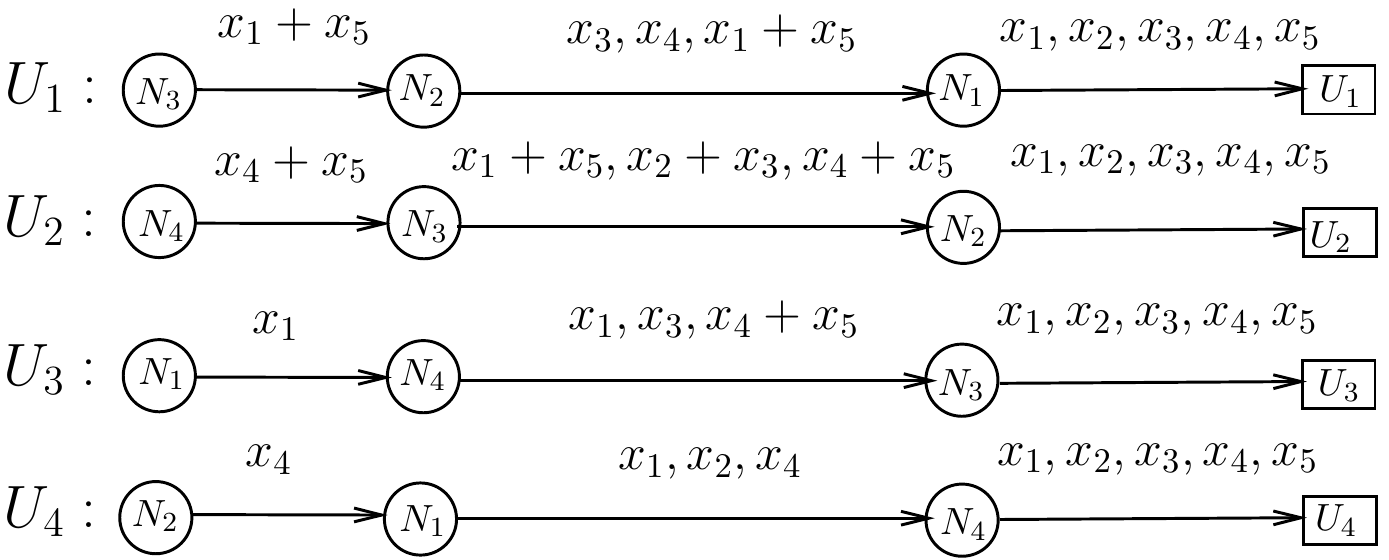}
\caption{The optimal reconstructing process of the new storage scheme.}
\label{fig_mrp}
\end{figure}

In fact, the above distributed storage scheme is not optimal. There exists a better storage scheme as described in Fig. \ref{fig_mrc}, where each user only
consumes reconstructing bandwidth 9 to recover the original data.
Fig. \ref{fig_mrp} characterizes its optimal reconstructing process.
Thus, the average minimum reconstructing bandwidth of this new storage scheme is 9. Actually, for this example, 9 is the minimum reconstructing bandwidth for any user.

\begin{figure}[!ht]
\centering
\includegraphics[width=11cm,height=4.3cm]{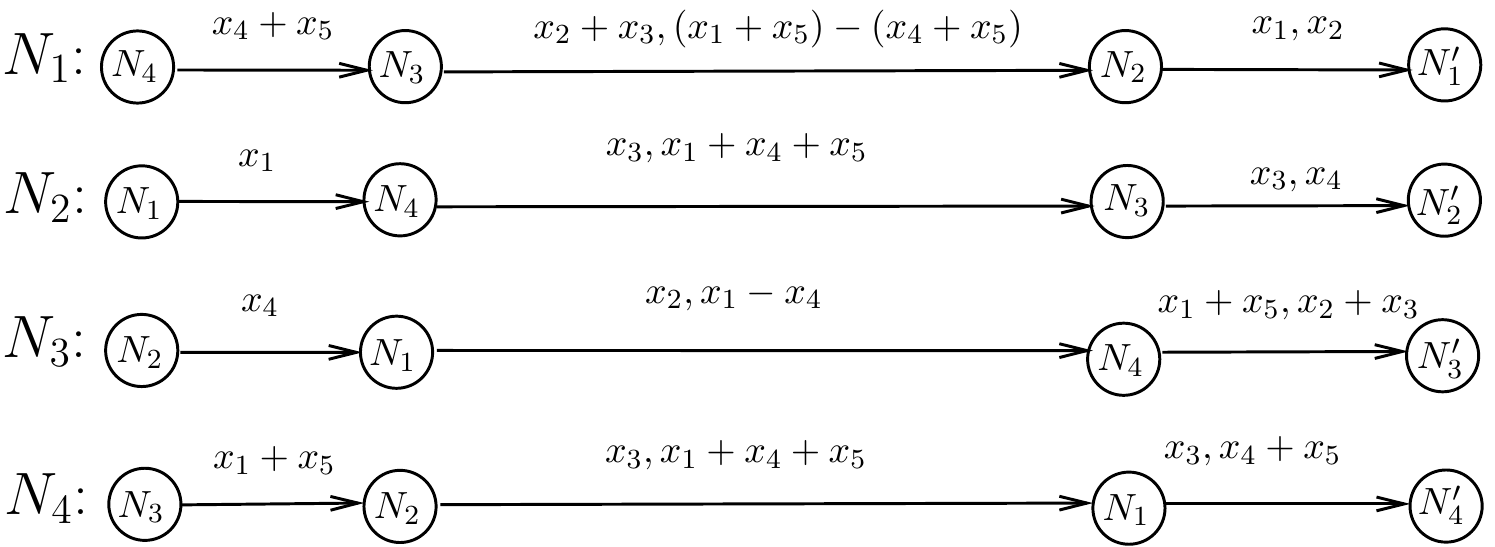}
\caption{The optimal repair process of the new storage scheme, where the $N_i^\prime (i=1,2,3,4),$ are the substituted nodes of the failed storage nodes.}
\label{fig_mrpair}
\end{figure}

Further, if some storage node fails, several interesting and meaningful problems should be considered. For instances, can this failed storage node be repaired? what is the minimum repair bandwidth? how to repair it efficiently? Fig. \ref{fig_mrpair} characterizes the optimal repair process with the minimum repair bandwidth 5 for each storage node of the new storage scheme depicted in Fig. \ref{fig_mrc}.
\subsection{Basic Model}\label{BM}
Let $\mathcal {G}$ be a unidirectional ring network consisting of $n$ storage nodes, denoted by $N_1,N_2,\cdots, N_n$. Each storage node has a capacity to store
$\alpha$ symbols. These $n$ storage nodes form a directed ring and data is only transmitted along the given direction. Let $X=[x_1,x_2,\cdots,x_M]$ be the row vector of original data with size $M$, each coordinate of which represents an information symbol taking values in a finite field $\mathbb{F}_q$ with $q$ elements, where $q$ is a power of some prime. The original data $X$ is distributed to all storage nodes in order to store $X$. Here, we just consider linear storage and linear transmission, that is, every stored symbol and every transmitted symbol are linear combinations of the information symbols, which are also elements in $\mathbb{F}_q$.

For any storage node $N_i$, define an $M \times \alpha$ node generator matrix $G^{(i)}$ over $\mathbb{F}_q$. Then all the $\alpha$ coordinates of the product $X G^{(i)}$ are stored in $N_i$, and each of which is called a node symbol. Each node symbol corresponds to a column vector of $G^{(i)}$, called a node vector. Further, each transmitted symbol is a linear combination of some node symbols. Clearly, it also corresponds to a vector, called a transmitted vector, which is the linear combination of those corresponding node vectors. Concatenating all node generator matrices according to the order of storage nodes, we obtain an $M \times n \alpha$ matrix  $G=[G^{(1)},G^{(2)},\cdots,G^{(n)}]$, which is called a generator matrix of a distributed storage scheme. Each user connects one and only one storage node to download data. Note that, in order to ensure that all users can reconstruct the original data completely in this ring network, the generator matrix $G$ has to be full row-rank.

We apply an information flow graph to analyze the reconstructing bandwidth, which is a particular graphical representation of distributed storage systems.

\begin{figure}[!ht]
\centering
\includegraphics[width=9cm,height=5.5cm]{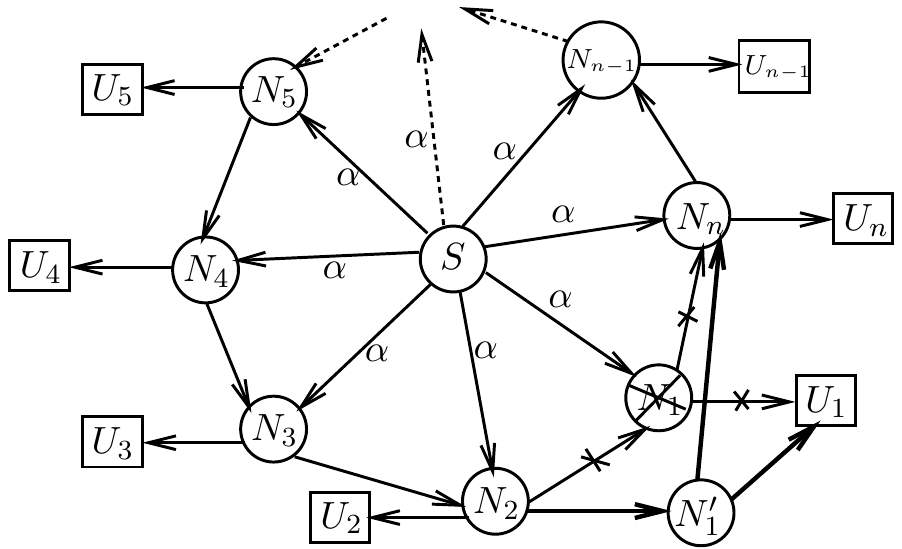}
\caption{The information flow graph $\mathcal {G}$, where circular nodes and rectangular nodes represent storage nodes and user nodes, respectively. When the storage node $N_1$ fails, a new node $N^\prime_{1}$ arises and establishes connections from the node $N_2$, to the node $N_n$ and the user node $U_1$.}
\label{inf-graph}
\end{figure}

$Information-Flow-Graph$: An information flow graph $\mathcal {G}$ consists of three types of nodes: a single source node $S$ (the source of original data $X$), $n$ storage nodes and some user nodes. The source node $S$ connects to $n$ storage nodes with directed edges of capacity $\alpha$. After the source node $S$ distributes node symbols, it becomes inactive. All storage nodes form a directed ring through edges with capacity $M$. The edges between user nodes and their corresponding storage nodes also have capacity $M$. When some storage node $N_i$ fails, a new substituted node $N^\prime_{i}$ arises to replace it and establishes connections from the node $N_{i+1}$, to the node $N_{i-1}$ and the users.  Due to the symmetry, we just take one user connecting to the same  storage node into account. Fig. \ref{inf-graph} indicates the details of the information flow graph $\mathcal {G}$.

In the following, we introduce some concepts in graph theory which are used in this paper.
A cut in the graph $\mathcal {G}$ between the source node $S$ and a fixed user $U$ is a subset of edges whose removal disconnects $S$ from $U$. The minimum cut  between $S$ and $U$ is a cut between them in which the total sum of the edge capacities achieves the smallest.

\section{Bounds on the Reconstructing and Repair Bandwidths}\label{B-RRB}
Recall that for unidirectional ring networks, there are three important parameters: $n$, the number of storage nodes; $\alpha$, the storage capacity per storage node; and $M$, the size of original data. In order to ensure entire storage of the original data, we have $n\geq \lceil M/\alpha \rceil$. By the ring structure, each user can always reconstruct the original data by connecting only one storage node. In this section, we mainly discuss the reconstructing bandwidth for each user and the repair bandwidth for each failed storage node.

\subsection{Lower Bound on the Reconstructing Bandwidth}
In the following, we first consider the reconstructing bandwidth for each user to recover the original data.

\begin{thm}\label{thm-rec-bound}
For any storage scheme of a unidirectional ring network with parameters $(n,\alpha,M)$, the reconstructing bandwidth for each user to recover the original data is lower bounded by $kM-\frac{(k-1)k\alpha}{2},$ where $k=\lceil M/\alpha \rceil$. Moreover, there exists a storage scheme such that all users can reconstruct the original data with reconstructing bandwidth achieving this lower bound with equality.
\end{thm}

Before the proof of Theorem \ref{thm-rec-bound}, we need the following two lemmas firstly.
\begin{lemma}\label{lemma-min-cut}{\cite[Lemma 1]{Dimakis-etc-2010}}
No user $U$ can reconstruct the original data if the minimum cut capacity between the source node $S$ and $U$ in a directed acyclic graph is smaller than the original data size $M$.
\end{lemma}

\begin{lemma}\label{lemma-iff}
For a storage scheme over a unidirectional ring network with parameters $(n,\alpha,M)$, all users can recover the original data with the same reconstructing bandwidth $kM-\frac{(k-1)k\alpha}{2}$, where $k=\lceil M/\alpha \rceil$, if and only if the following two conditions are satisfied:

(i) all $(k-1)\alpha$ node vectors of arbitrary $k-1$ adjacent storage nodes are linearly independent;

(ii) arbitrary $k$ adjacent storage nodes contain $M$ linearly independent node vectors.
\end{lemma}

To keep the continuity of the paper, the complete proof of this lemma is given in Appendix \ref{app1}.

\begin{IEEEproof}[Proof of Theorem \ref{thm-rec-bound}]
First, we discuss the reconstructing bandwidth for each user. Notice that,
because of the symmetry of this network, it suffices to consider the reconstructing bandwidth for any one user. Without loss of generality, we take the user $U_1$ into account, which is connected from the storage node $N_1$ as depicted in Fig. \ref{inf-graph}. Clearly, when $U_1$ is under consideration, it is not necessary to transmit data from $N_1$ to $N_n$. So we can omit the edge from $N_1$ to $N_n$ and only consider the degenerated acyclic graph $\mathcal {H}$ of the cyclic graph $\mathcal {G}$ as depicted in Fig. \ref{fig_ag}.
\begin{figure}[!ht]
\centering
\includegraphics[width=9cm,height=5.5cm]{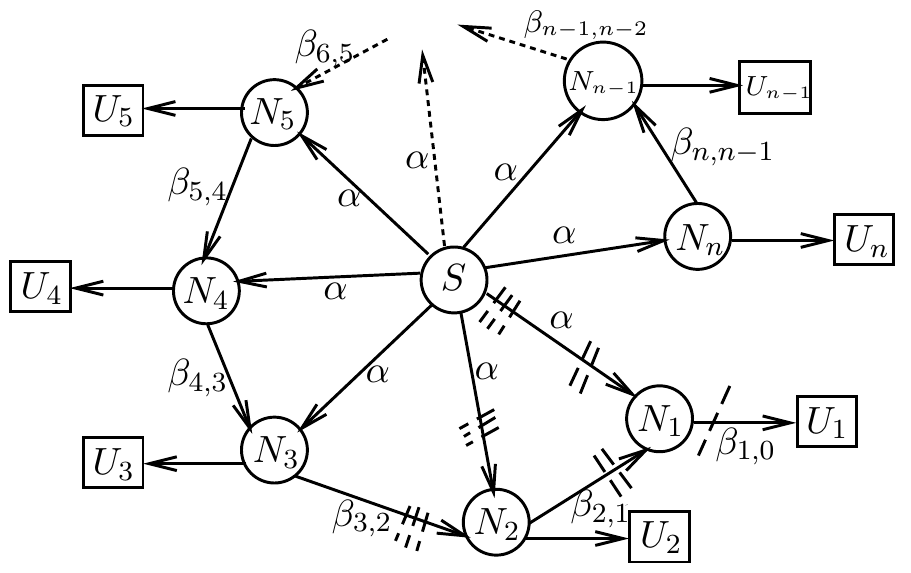}
\caption{The degenerated acyclic graph $\mathcal {H}$. All storage nodes form a chain for $U_1$ to reconstruct the original data.}
\label{fig_ag}
\end{figure}

Moreover, in the acyclic graph $\mathcal {H}$, let $\beta_{i,i-1}$ be the number of transmitted vectors (equivalently, transmitted symbols) from the storage node $N_i$ to the storage node $N_{i-1}$, where $1\leq i \leq n$ and $N_0$ represents the user node $U_1$ in order to keep consistency of notation. Thus, the total reconstructing bandwidth for the user $U_1$ is $\sum^{n}_{i=1}\beta_{i,i-1}$. In the following, we will propose a lower bound by cut-set bound analysis method in Fig. \ref{fig_ag}.

In order to ensure the user $U_1$ to reconstruct the original data,
by Lemma \ref{lemma-min-cut}, the minimum cut capacity between the source node $S$ and the user node $U_1$ should not be less than $M$. This implies that we just need to analyze those potential minimum cuts between $S$ and $U_1$. For example, as depicted in Fig. \ref{fig_ag}, the link $(N_1,U_1)$ is a potential minimum cut between $S$ and $U_1$, so it has to satisfy $\beta_{1,0}\geq M$. For another example, the set of links $\{(S,N_1),(N_2,N_1)\}$ is another potential minimum cut between $S$ and $U_1$, so it has to satisfy $\alpha+\beta_{2,1}\geq M$. In the following, we list inequalities for all potential minimum cuts between $S$ and $U_1$ as follows:
\begin{equation*}
\begin{aligned}
\beta_{1,0} & \geq M, \\
\alpha+\beta_{2,1} & \geq M, \\
\vdots\\
(k-1)\alpha+\beta_{k,k-1} & \geq M, \\
k\alpha+\beta_{k+1,k} & \geq M,\\
\vdots \\
(n-1)\alpha+\beta_{n,n-1} & \geq M,
\end{aligned}
\end{equation*}
where again $k=\lceil M/\alpha \rceil$.
Note that we have $k\alpha\geq M$ as $k=\lceil M/\alpha\rceil$, which implies that the last $(n-k)$ inequalities are useless. By the first $k$ inequalities above, we can obtain the lower bounds on $\beta_{i,i-1}$ for $1\leq i \leq k$, that is,
\begin{equation*}
\begin{aligned}
\beta_{1,0} & \geq M, \\
\beta_{2,1} & \geq M-\alpha, \\
\vdots\\
\beta_{k,k-1} & \geq M-(k-1)\alpha.
\end{aligned}
\end{equation*}
Hence, the above analysis implies the lower bound below:
$$\sum^{n}_{i=1}\beta_{i,i-1}\geq M+(M-\alpha)+\cdots+[M-(k-1)\alpha]=kM-\frac{(k-1)k\alpha}{2}.$$

Next, we further indicate the tightness of this bound. We will construct a storage scheme such that all users can recover the original data with the same reconstructing bandwidth $kM-\frac{(k-1)k\alpha}{2}$.

Let $\mathbb{F}_q$ be the based field with $q$ elements, where $q$ is a power of some prime. Let $G$ represent the generator matrix of a distributed storage scheme. Note that $G$ is an $M\times n\alpha$ matrix over $\mathbb{F}_q$.  Denote $g_i$ the $i$th column vector of $G$ for $1\leq i\leq n\alpha$.
Now, we will construct a generator matrix $G$ to obtain a proper distributed storage scheme. First, we easily select $M$ linearly independent vectors $g_1,g_2,\cdots,g_M$ from the vector space $\mathbb{F}^M_q$. Next, we will use the following procedure to choose the remaining $(n\alpha-M)$ column vectors of $G$.  We construct the $i$th column vector $g_i$ for $M+1\leq i \leq n\alpha$ successively such that $g_i$ is linearly independent with previous arbitrary $(M-1)$ column vectors.

For each integer $i$ with $M+1\leq i \leq n\alpha$, let $[i-1]\triangleq\{1,2,\cdots,i-1\}$ and $\mathcal {N}_{i-1}\triangleq\{I\subset[i-1]:|I|=M-1\}$, where $|I|$ denotes the size of $I$. Then, we select $g_i$ such that the following is satisfied:
$$g_{i}\in \mathbb{F}^M_q \backslash \bigcup\limits_{I\in \mathcal {N}_{i-1}}\langle g_j:j\in I \rangle,$$
where $\langle g_j: j\in I\rangle $ represents the vector space spanned by the vectors $g_j$, $j\in I$.
Furthermore, it is not difficult to see that $g_i$ can be always chosen for each $i$, $M+1\leq i \leq n\alpha$, if the size of the finite field $\mathbb{F}_q$ is sufficiently large. Actually, it is enough that $|\mathbb{F}_q|>|\mathcal {N}_{n\alpha-1}|=\binom{n\alpha-1}{M-1}$.
Therefore, we can construct a generator matrix $G=[g_1,g_2,\cdots,g_{n\alpha}]$ satisfying that its arbitrary $M$ column vectors are linearly independent. Then we partition all $n\alpha$ column vectors of $G$ into $n$ parts, each of which contains $\alpha$ column vectors constituting the node generator matrix of a storage node. Thus, we obtain a storage scheme. Particularly, it is easy to check that this storage scheme satisfies the two conditions in Lemma \ref{lemma-iff}. Therefore, all users can recover the original data with the same reconstructing bandwidth $kM-\frac{(k-1)k\alpha}{2}$. This completes the proof.
\end{IEEEproof}
\begin{remark}

(\rom 1) In order to reconstruct the original data for each user, $k$ is actually the minimum number of storage nodes which need to transmit data.

(\rom 2) If a distributed storage scheme achieves the lower bound in Theorem \ref{thm-rec-bound} with equality for all users, we say it an optimal reconstructing distributed storage scheme (ORDSS). Actually, the second part of the proof of Theorem \ref{thm-rec-bound} implies a construction method of ORDSSes, which is also an approach for constructing a generator matrix of an $[n\alpha, M]$ maximum distance separable (MDS) code in algebraic coding theory \cite{M-S}. Inversely, a generator matrix of any $[n\alpha, M]$ MDS can be used as a generator matrix of an ORDSS. Thus, we say this construction method MDS construction. In addition, by the above definition of an ORDSS, the two conditions in Lemma \ref{lemma-iff} are actually sufficient and necessary for the existence of an ORDSS.
\end{remark}

\begin{eg}\label{EX_2}
For a unidirectional ring network with parameters $(n=4,\alpha=2,M=5)$, let the finite field be $\mathbb{F}_{11}$ and the row vector of original data be $X=[x_1,x_2,x_3,x_4,x_5]\in \mathbb{F}_{11}^5$. We select a generator matrix $G$ of an $[8,5]$ MDS code over $\mathbb{F}_{11}$ as follows:
$$G=\left[\begin{array}{cccccccc}1&0&0&0&0&1&5&4\\0&1&0&0&0&6&9&7\\0&0&1&0&0&10&1&5\\0&0&0&1&0&5&4&2\\0&0&0&0&1&1&4&5\end{array}\right].$$
Subsequently, we have
$$(XG)^\top =\left[\begin{array}{cccccccc}x_1\\x_2\\x_3\\x_4\\x_5\\x_1+6x_2+10x_3+5x_4+x_5\\5x_1+9x_2+x_3+4x_4+4x_5\\4x_1+7x_2+5x_3+2x_4+5x_5
\end{array}\right],$$
where $(XG)^\top$ represents the transposition of $XG$. Then we distribute the eight coordinates of $XG$ to the four storage nodes arbitrarily, each of which stores two coordinates. One of storage schemes is depicted in Fig. \ref{fig_ex2}, and Fig. \ref{fig_ex2_u1} describes an optimal reconstructing process of the user $U_1$.  We can analyze the other users similarly.
\begin{figure}[!ht]
\centering
\includegraphics[width=7.5cm,height=5.2cm]{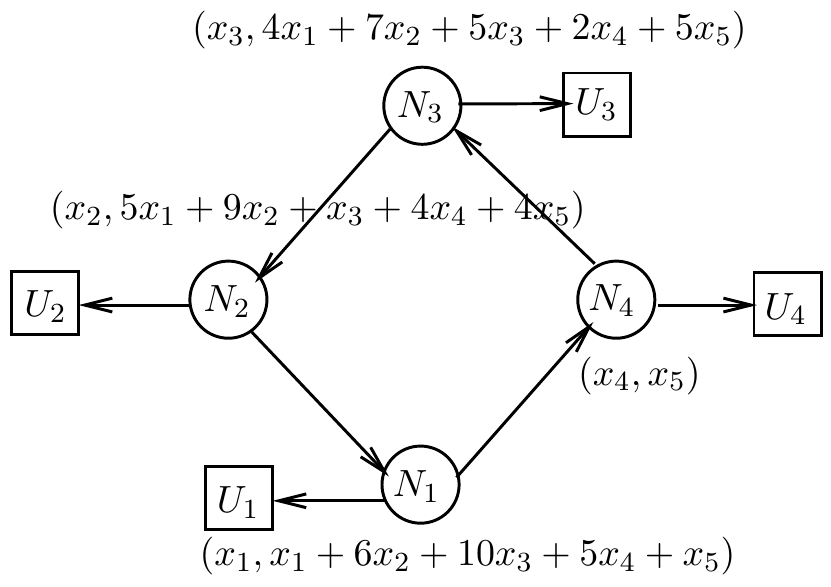}
\caption{The storage scheme with a generator matrix of an MDS code as its generator matrix.}
\label{fig_ex2}
\end{figure}
\begin{figure}[!ht]
\centering
\includegraphics[width=11cm,height=1cm]{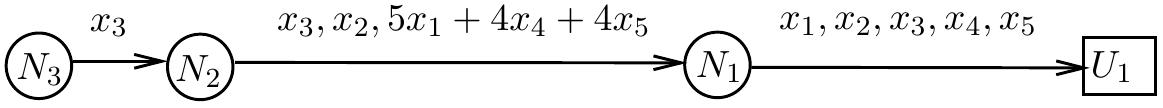}
\caption{The reconstructing process for $U_1$. $N_2$ receives $x_3$ and eliminates $9x_2+x_3$ from $5x_1+9x_2+x_3+4x_4+4x_5$.}
\label{fig_ex2_u1}
\end{figure}
In Fig. \ref{fig_ex2_u1}, $N_3$ transmits one of its own node symbol $x_3$ to $N_2$, after a simple calculation, $N_2$ transmits three symbols $x_3$, $x_2$ and $5x_1+4x_4+4x_5$ to $N_1$, then $N_1$ calculates the original data $(x_1,x_2,x_3,x_4,x_5)$ by combining the three received symbols with its own two node symbols, and finally outputs to the user $U_1$. Clearly, the reconstructing bandwidth for the user $U_1$ to recover the original data is 9, equal to the proposed lower bound $kM-\frac{(k-1)k\alpha}{2}$ with $\alpha=2,M=5,k=\lceil M/\alpha \rceil=3$. Actually, all users can recover the original data with reconstructing bandwidth 9. So this scheme is an ORDSS.
\end{eg}

\subsection{Lower Bound on the Repair Bandwidth}
In an ORDSS, when a storage node fails, a new node arises to replace it. By some data transmission over the network, if the data stored in this new node keeps the same as that in the original node before failing, then the failed node is said to be repaired successfully. The total number of transmitted symbols for repairing this failed node is called repair bandwidth. In this subsection, we will focus on the repair bandwidth for any failed storage node in any ORDSS. First, we give a lower bound on the repair bandwidth. Different from the proof of Theorem \ref{thm-rec-bound}, here, we mainly apply linear algebra as analysis technique.
\begin{thm}\label{thm-repair}
For any ORDSS over a unidirectional ring network with parameters $(n,\alpha,M)$, the repair bandwidth for any failed storage node is lower bounded by $M$.
\end{thm}

\begin{IEEEproof}
We only take the repair of the storage node $N_1$ into account as the symmetry of the network. In order to repair the failed node $N_1$, a new node $N^\prime_{1}$ arises to replace it and establishes connections from the storage node $N_2$, to the storage node $N_n$ and the user node $U_1$. Furthermore, the new node $N^\prime_{1}$ and the other remaining storage nodes still form a unidirectional ring. Let $M=(k-1)\alpha+\gamma$, where $k=\lceil M/\alpha \rceil$ and $0<\gamma\leq\alpha$.

According to Lemma \ref{lemma-iff}, for an ORDSS, all $(k-1)\alpha$ node vectors of arbitrary $k-1$ adjacent storage nodes are linearly independent and  arbitrary $k$ adjacent storage nodes contain $M$ linearly independent node vectors. Particularly, denote all $\alpha$ linearly independent node vectors in $N_i$ by $\textbf{a}_{i}^{1},\textbf{a}_{i}^{2},\cdots,\textbf{a}_{i}^{\alpha}$, $1\leq i\leq n$. Let $\xi_{i,i-1}$ be the number of transmitted vectors from $N_i$ to $N_{i-1}$, $3\leq i\leq n$, and $\xi_{2,1}$ be the number of transmitted vectors from $N_2$ to $N^\prime_{1}$ for repairing $N_1$. We discuss $\xi_{2,1}$ first. Since the storage node $N_1$ consists of $\alpha$ linearly independent node vectors, it is impossible to be repaired by less than $\alpha$ vectors. Thus, we deduce $\xi_{2,1}\geq\alpha$.

Next, we consider $\xi_{3,2}$, the number of transmitted vectors from $N_3$ to $N_2$. We claim $\xi_{3,2}\geq\alpha$. Conversely, assume $\xi_{3,2}<\alpha$ and let the $\xi_{3,2}$ transmitted vectors be $\textbf{b}_{3}^{1},\textbf{b}_{3}^{2},\cdots,\textbf{ b}_{3}^{\xi_{3,2}}$. If the storage node $N_1$ is repaired successfully, then all $\alpha$ node vectors in $N_1$ can be linearly expressed  by the $\alpha$ node vectors stored in $N_2$ and the $\xi_{3,2}$ transmitted vectors from $N_3$ to $N_2$, that is,
\begin{equation}
\begin{aligned}
\textbf{a}_{1}^{1}&=\sum^{\xi_{3,2}}_{i=1}s_{1,i}\textbf{b}_{3}^{i}+\sum^{\alpha}_{i=1}t_{1,i}\textbf{a}_{2}^{i},\\
\vdots\\
\textbf{a}_{1}^{\alpha}&=\sum^{\xi_{3,2}}_{i=1}s_{\alpha,i}\textbf{b}_{3}^{i}+\sum^{\alpha}_{i=1}t_{\alpha,i}\textbf{a}_{2}^{i}.
\end{aligned}
\end{equation}
where $s_{1,i},\cdots,s_{\alpha,i} \in \mathbb{F}_q $ for $1\leq i\leq \xi_{3,2}$ and $t_{1,i},\cdots,t_{\alpha,i} \in \mathbb{F}_q $ for $1\leq i\leq \alpha$.
Since the $\alpha$ vectors $\sum^{\xi_{3,2}}_{i=1}s_{1,i}\textbf{b}_{3}^{i},\cdots,\sum^{\xi_{3,2}}_{i=1}s_{\alpha,i}\textbf{b}_{3}^{i}$ are linear combinations of the $\xi_{3,2}$ vectors $\textbf{b}_{3}^{1},\cdots,\textbf{b}_{3}^{\xi_{3,2}}$, together with $\xi_{3,2}<\alpha$, it follows that $\sum^{\xi_{3,2}}_{i=1}s_{j,i}\textbf{b}_{3}^{i}$, $1\leq j\leq\alpha$, are linearly dependent. Thus, there exists at least one of them which can be expressed linearly by the others. Without loss of generality, assume that
$$\sum^{\xi_{3,2}}_{i=1}s_{\alpha,i}\textbf{b}_{3}^{i}=l_1 \sum^{\xi_{3,2}}_{i=1}s_{1,i}\textbf{b}_{3}^{i}+l_2 \sum^{\xi_{3,2}}_{i=1}s_{2,i}\textbf{b}_{3}^{i}+\cdots+l_{\alpha-1}\sum^{\xi_{3,2}}_{i=1}s_{\alpha-1,i}\textbf{b}_{3}^{i},$$ where $l_1,l_2,\cdots,l_{\alpha-1}\in\mathbb{F}_q$.
Combining with the equations in (1), we have
\begin{align*}
\textbf{a}_{1}^{\alpha} & = l_1 \sum^{\xi_{3,2}}_{i=1}s_{1,i}\textbf{b}_{3}^{i}+l_2 \sum^{\xi_{3,2}}_{i=1}s_{2,i}\textbf{b}_{3}^{i}+\cdots+l_{\alpha-1}\sum^{\xi_{3,2}}_{i=1}s_{\alpha-1,i}\textbf{b}_{3}^{i}
+\sum^{\alpha}_{i=1}t_{\alpha,i}\textbf{a}_{2}^{i}\\
       & = l_1(\textbf{a}_{1}^{1}-\sum^{\alpha}_{i=1}t_{1,i}\textbf{a}_{2}^{i})+\cdots+l_{\alpha-1}(\textbf{a}_{1}^{\alpha-1}
-\sum^{\alpha}_{i=1}t_{\alpha-1,i}\textbf{a}_{2}^{i})+\sum^{\alpha}_{i=1}t_{\alpha,i}\textbf{a}_{2}^{i}\\
       & = l_1\textbf{a}_{1}^{1}+\cdots+l_{\alpha-1}\textbf{a}_{1}^{\alpha-1}+\sum^{\alpha}_{i=1}t_{\alpha,i}\textbf{a}_{2}^{i}-l_1 \sum^{\alpha}_{i=1}t_{1,i}\textbf{a}_{2}^{i}-\cdots-l_{\alpha-1}\sum^{\alpha}_{i=1}t_{\alpha-1,i}\textbf{a}_{2}^{i}.
\end{align*}
This implies that $\textbf{a}_{1}^{1},\cdots,\textbf{a}_{1}^{\alpha},\textbf{a}_{2}^{1},\cdots,\textbf{a}_{2}^{\alpha}$ are linearly dependent, which leads to a contradiction. Hence, we obtain $\xi_{3,2} \geq \alpha$.

Continuing this process and using the similar analysis, it is easy to see that $\xi_{i,i-1}$ is not less than $\alpha$ for $2\leq i\leq k$, that is, $\xi_{i,i-1}\geq \alpha$ for $2\leq i\leq k$.

At last, we discuss the transmission process from $N_{k+1}$ to $N_k$. Recall that $\xi_{k+1,k}$ represents the number of transmitted vectors from $N_{k+1}$ to $N_k$. For notation simplicity, let $d \triangleq \xi_{k+1,k}$ and we claim $d\geq\gamma$, where $\gamma=M-(k-1)\alpha$. Assume the contrary that $d<\gamma$ and denote the $d$ transmitted vectors by $\textbf{b}_{k+1}^{1},\cdots,\textbf{b}_{k+1}^{d}$. The storage node $N_1$ can be repaired successfully, so the $\alpha$ node vectors stored in $N_1$ can be expressed linearly by the $(k-1)\alpha$ node vectors in $N_2,\cdots,N_k$ and the $d$ transmitted vectors $\textbf{b}_{k+1}^{1},\cdots,\textbf{b}_{k+1}^{d}$ from $N_{k+1}$ to $N_k$. Thus we have
\begin{equation}
\begin{aligned}
\textbf{a}_{1}^{1}&=\sum^d_{i=1}s_{1,i}\textbf{b}_{k+1}^{i}+\sum^{\alpha}_{j=1}t^{(2)}_{1,j}\textbf{a}_{2}^{j}+\cdots+\sum^{\alpha}_{j=1}t^{(k)}_{1,j}
\textbf{a}_{k}^{j},\\
\vdots\\
\textbf{a}_{1}^{\alpha}&=\sum^d_{i=1}s_{\alpha,i}\textbf{b}_{k+1}^{i}+\sum^{\alpha}_{j=1}t^{(2)}_{\alpha,j}\textbf{a}_{2}^{j}+\cdots+\sum^{\alpha}_{j=1}
t^{(k)}_{\alpha,j}\textbf{a}_{k}^{j},
\end{aligned}
\end{equation}
where all coefficients $s_{1,i},s_{2,i},\cdots,s_{\alpha,i}$ and $t^{(l)}_{1,j},t^{(l)}_{2,j},\cdots,t^{(l)}_{\alpha,j}$, $1\leq i\leq d,1\leq j\leq\alpha,2\leq l\leq k$, are elements in $\mathbb{F}_q$.
Consider the $\alpha$ vectors $\sum^d_{i=1}s_{1,i}\textbf{b}_{k+1}^{i}, \sum^d_{i=1}s_{2,i}\textbf{b}_{k+1}^{i}, \cdots,\sum^d_{i=1}s_{\alpha,i}\textbf{b}_{k+1}^{i}$,
 in which the maximum number of linearly independent vectors is not greater than $d$. This implies that there exist at least $(\alpha-d)$ vectors, which can be expressed linearly by the others. Without loss of generality, assume that the first $(\alpha-d)$ vectors $\sum^d_{i=1}s_{1,i}\textbf{b}_{k+1}^{i},\sum^d_{i=1}s_{2,i}\textbf{b}_{k+1}^{i}, \cdots,\sum^d_{i=1}s_{\alpha-d,i}\textbf{b}_{k+1}^{i}$ can be represented linearly by the other $d$ vectors. Thus, we have
\begin{equation}
\begin{aligned}
\sum^d_{i=1}s_{1,i}\textbf{b}_{k+1}^{i}&= l_{1,1}\sum^d_{i=1}s_{\alpha-d+1,i}\textbf{b}_{k+1}^{i}+\cdots+l_{1,d}\sum^d_{i=1}s_{\alpha,i}\textbf{b}_{k+1}^{i},\\
\vdots\\
\sum^d_{i=1}s_{\alpha-d,i}\textbf{b}_{k+1}^{i}&= l_{\alpha-d,1}\sum^d_{i=1}s_{\alpha-d+1,i}\textbf{b}_{k+1}^{i}+\cdots+l_{\alpha-d,d}\sum^d_{i=1}s_{\alpha, i}\textbf{b}_{k+1}^{i},
\end{aligned}
\end{equation}
where the coefficients $l_{i,j}$, $1\leq i\leq\alpha-d,1\leq j\leq d$ are still elements in $\mathbb{F}_q$.

Combining (2) and (3), for any $1\leq r\leq \alpha-d$, we obtain that $\textbf{a}_{1}^{r}$ can be expressed as follows£º
\begin{equation}
\begin{aligned}
\textbf{a}_{1}^{r}&=l_{r,1}\sum^d_{i=1}s_{\alpha-d+1,i}\textbf{b}_{k+1}^{i}+\cdots+l_{r,d}\sum^d_{i=1}s_{\alpha,i}\textbf{b}_{k+1}^{i}
+\sum^{\alpha}_{j=1}t^{(2)}_{r,j}\textbf{a}_{2}^{j}+\cdots+\sum^{\alpha}_{j=1}t^{(k)}_{r,j}\textbf{a}_{k}^{j}\\
&=l_{r,1}(\textbf{a}_{1}^{\alpha-d+1}-\sum^{\alpha}_{j=1}t^{(2)}_{\alpha-d+1,j}\textbf{a}_{2}^{j}
-\cdots-\sum^{\alpha}_{j=1}t^{(k)}_{\alpha-d+1,j}\textbf{a}_{k}^{j})+\cdots\\
&\quad+l_{r,d}(\textbf{a}_{1}^{\alpha}-\sum^{\alpha}_{j=1}t^{(2)}_{\alpha,j}\textbf{a}_{2}^{j}-\cdots-\sum^{\alpha}_{j=1}t^{(k)}_{\alpha,j}\textbf{a}_{k}^{j})
+\sum^{\alpha}_{j=1}t^{(2)}_{r,j}\textbf{a}_{2}^{j}+\cdots+\sum^{\alpha}_{j=1}t^{(k)}_{r,j}\textbf{a}_{k}^{j}\\
&=l_{r,1}\textbf{a}_{1}^{\alpha-d+1}+\cdots+l_{r,d}\textbf{a}_{1}^{\alpha}\\
&\quad+(\sum^{\alpha}_{j=1}t^{(2)}_{r,j}\textbf{a}_{2}^{j}-l_{r,1}\sum^{\alpha}_{j=1}t^{(2)}_{\alpha-d+1,j}\textbf{a}_{2}^{j}
-\cdots-l_{r,d}\sum^{\alpha}_{j=1}t^{(2)}_{\alpha,j}\textbf{a}_{2}^{j})+\cdots\\
&\quad+(\sum^{\alpha}_{j=1}t^{(k)}_{r,j}\textbf{a}_{k}^{j}-l_{r,1}\sum^{\alpha}_{j=1}t^{(k)}_{\alpha-d+1,j}\textbf{a}_{k}^{j}
-\cdots-l_{r,d}\sum^{\alpha}_{j=1}t^{(k)}_{\alpha,j}\textbf{a}_{k}^{j}).\\
\end{aligned}
\end{equation}
The equations in (4) indicate that the $\alpha-d$ node vectors $\textbf{a}_{1}^{1},\textbf{a}_{1}^{2},\cdots,\textbf{a}_{1}^{\alpha-d}$ are linear combinations of the $d$ node vectors $\textbf{a}_{1}^{\alpha-d+1},\cdots,\textbf{a}_{1}^{\alpha}$, stored in $N_{1}$, and all $(k-1)\alpha$ node vectors, stored in $N_2,N_3,\cdots,N_k$. Thus, the number of linearly independent vectors stored in $N_1,N_2,\cdots,N_k$ is not greater than $(k-1)\alpha+d<(k-1)\alpha+\gamma=M$. This contradicts to the property (\rom 2) of the ORDSS in Lemma \ref{lemma-iff}. Hence, $\xi_{k+1,k}\geq\gamma$.

Until now, $N_2,N_3,\cdots,N_{k+1}$ have to provide at least $\sum^{k+1}_{i=2}\xi_{i,i-1}=(k-1)\alpha+\gamma=M$ vectors to repair $N_1$. Thus, $M$ is a lower bound on the repair bandwidth, which accomplishes the proof.
\end{IEEEproof}

For the above lower bound on the repair bandwidth, we hope to know whether there exists an ORDSS achieving this bound with equality for each storage node. The following theorem answers this question.

\begin{thm}\label{thm-ORDSS-repair}
For any ORDSS over a unidirectional ring network with parameters $(n,\alpha,M)$, every storage node can be repaired successfully with the repair bandwidth $M$ if it fails.
\end{thm}

\begin{IEEEproof}
Similarly, for any ORDSS, it is sufficient to discuss any one storage node because of the symmetry of the network and we still take the storage node $N_1$ into account. Furthermore, we just need to verify that $N_1$ can be repaired with the repair bandwidth $M$. We still let $\textbf{a}_{i}^{1},\textbf{a}_{i}^{2},\cdots,\textbf{a}_{i}^{\alpha}$ represent the $\alpha$ linearly independent node vectors in the storage node $N_i$ for $1\leq i\leq n$.

Note that $N_1,N_2,\cdots,N_k$ are $k$ adjacent storage nodes. By Lemma \ref{lemma-iff}, for the ORDSS, all $(k-1)\alpha$ node vectors in $N_1,N_2,\cdots,N_{k-1}$ are linearly independent and there are $M$ linearly independent node vectors in $N_1,N_2,\cdots,N_k$. Without loss of generality, we assume that $\textbf{a}_{i}^{1},\textbf{a}_{i}^{2},\cdots,\textbf{a}_{i}^{\alpha}$ in $N_{i}$, $1\leq i\leq k-1$, and $\textbf{a}_{k}^{1},\textbf{a}_{k}^{2},\cdots,\textbf{a}_{k}^{\gamma}$ in $N_{k}$ are $M$ linearly independent node vectors. Thus, the remaining node vectors can be expressed linearly by these $M$ vectors.

Since $N_{2},N_{3}\cdots,N_{k+1}$ are another $k$ adjacent storage nodes in the ORDSS. Similarly, we also assume that $\textbf{a}_{i}^{1},\textbf{a}_{i}^{2},\cdots,\textbf{a}_{i}^{\alpha}$ in $N_i$, $2\leq i\leq k$, and  $\textbf{a}_{k+1}^{1},\textbf{a}_{k+1}^{2},\cdots,\textbf{a}_{k+1}^{\gamma}$ in $N_{k+1}$, are $M$ linearly independent node vectors. Moreover, we can use $\textbf{a}_{1}^{1},\cdots,\textbf{a}_{1}^{\alpha},\cdots,$ $\textbf{a}_{k-1}^{1}\cdots,\textbf{a}_{k-1}^{\alpha},\textbf{a}_{k}^{1},\cdots,\textbf{a}_{k}^{\gamma}$ to express the $\alpha$ node vectors $\textbf{a}_{k}^{\gamma+1}\cdots,\textbf{a}_{k}^{\alpha}$, $\textbf{a}_{k+1}^{1},\cdots,\textbf{a}_{k+1}^{\gamma}$ linearly, specifically,
\begin{equation}
\begin{aligned}
\textbf{a}_{k}^{u}&=\sum^\alpha_{i=1}c^{(1)}_{u,i}\textbf{a}_{1}^{i}+\sum^\alpha_{i=1}c^{(2)}_{u,i}\textbf{a}_{2}^{i}+\cdots
+\sum^\alpha_{i=1}c^{(k-1)}_{u,i}\textbf{a}_{k-1}^{i}+\sum^\gamma_{i=1}c^{(k)}_{u,i}\textbf{a}_{k}^{i}\quad(\gamma+1\leq u \leq \alpha),\\
\textbf{a}_{k+1}^{v}&=\sum^\alpha_{i=1}d^{(1)}_{v,i}\textbf{a}_{1}^{i}+\sum^\alpha_{i=1}d^{(2)}_{v,i}\textbf{a}_{2}^{i}+\cdots
+\sum^\alpha_{i=1}d^{(k-1)}_{v,i}\textbf{a}_{k-1}^{i}+\sum^\gamma_{i=1}d^{(k)}_{v,i}\textbf{a}_{k}^{i}\qquad(1\leq v \leq \gamma),
\end{aligned}
\end{equation}
 with all coefficients in $\mathbb{F}_q$.

Subsequently, we present the detailed repair process of the storage node $N_1$. In order to repair $N_1$, the storage node $N_{k+1}$ transmits the $\gamma$ node vectors $\textbf{a}_{k+1}^{1},\textbf{a}_{k+1}^{2},\cdots,\textbf{a}_{k+1}^{\gamma}$ to the storage node $N_k$, then $N_k$ eliminates the terms $\sum^\gamma_{i=1}c^{(k)}_{u,i}\textbf{a}_{k}^{i}$ and $\sum^\gamma_{i=1}d^{(k)}_{v,i}\textbf{a}_{k}^{i}$ from the above expressions of $\textbf{a}_{k}^{u}$ and $\textbf{a}_{k+1}^{v}$ for $\gamma+1\leq u\leq \alpha$, $1\leq v\leq \gamma$, and transmits the following $\alpha$ vectors to the storage node $N_{k-1}$:
\begin{equation}
\begin{aligned}
\sum^\alpha_{i=1}c^{(1)}_{u,i}\textbf{a}_{1}^{i}+\sum^\alpha_{i=1}c^{(2)}_{u,i}\textbf{a}_{2}^{i}+\cdots+\sum^\alpha_{i=1}c^{(k-1)}_{u,i}\textbf{a}_{k-1}^{i}
\qquad (\gamma+1\leq u\leq \alpha),\\
\sum^\alpha_{i=1}d^{(1)}_{v,i}\textbf{a}_{1}^{i}+\sum^\alpha_{i=1}d^{(2)}_{v,i}\textbf{a}_{2}^{i}+\cdots+\sum^\alpha_{i=1}d^{(k-1)}_{v,i}\textbf{a}_{k-1}^{i} \qquad \qquad (1\leq v\leq \gamma).
\end{aligned}
\end{equation}
Then, using its node vectors, $N_{k-1}$ eliminates the terms $\sum^\alpha_{i=1}c^{(k-1)}_{u,i}\textbf{a}_{k-1}^{i}$ and $\sum^\alpha_{i=1}d^{(k-1)}_{v,i}\textbf{a}_{k-1}^{i}$ from the received $\alpha$ vectors in (6) and transmits the $\alpha$ vectors
below to the storage node $N_{k-2}$:
$$\sum^\alpha_{i=1}c^{(1)}_{u,i}\textbf{a}_{1}^{i}+\sum^\alpha_{i=1}c^{(2)}_{u,i}\textbf{a}_{2}^{i}+\cdots+\sum^\alpha_{i=1}c^{(k-2)}_{u,i}\textbf{a}_{k-1}^{i}
\qquad (\gamma+1\leq u\leq \alpha),$$
$$\sum^\alpha_{i=1}d^{(1)}_{v,i}\textbf{a}_{1}^{i}+\sum^\alpha_{i=1}d^{(2)}_{v,i}\textbf{a}_{2}^{i}+\cdots+\sum^\alpha_{i=1}d^{(k-2)}_{v,i}\textbf{a}_{k-1}^{i} \qquad \qquad (1\leq v\leq \gamma).$$
Continuing the process until the storage node $N_2$, it receives the following $\alpha$ vectors:

$$\sum^\alpha_{i=1}c^{(1)}_{u,i}\textbf{a}_{1}^{i}+\sum^\alpha_{i=1}c^{(2)}_{u,i}\textbf{a}_{2}^{i}\qquad (\gamma+1\leq u\leq \alpha),$$
$$\sum^\alpha_{i=1}d^{(1)}_{v,i}\textbf{a}_{1}^{i}+\sum^\alpha_{i=1}d^{(2)}_{v,i}\textbf{a}_{2}^{i} \qquad \qquad (1\leq v\leq \gamma),$$
and uses its node vectors to eliminate the terms $\sum^\alpha_{i=1}c^{(2)}_{u,i}\textbf{a}_{2}^{i}$ and $\sum^\alpha_{i=1}d^{(2)}_{v,i}\textbf{a}_{2}^{i}$, $\gamma+1\leq u\leq \alpha,$ $1\leq v\leq \gamma$. Next, $N_2$ transmits the $\alpha$
vectors $\sum^\alpha_{i=1}c^{(1)}_{\gamma+1,i}\textbf{a}_{1}^{i},\cdots,\sum^\alpha_{i=1}c^{(1)}_{\alpha,i}\textbf{a}_{1}^{i}$, $\sum^\alpha_{i=1}d^{(1)}_{1,i}\textbf{a}_{1}^{i},\cdots,\sum^\alpha_{i=1}d^{(1)}_{\gamma,i}\textbf{a}_{1}^{i}$ to the new node $N_{1}^{\prime}$ for repairing $N_1$.

Finally, we claim that the $\alpha$ vectors transmitted to the new node $N_{1}^{\prime}$ are linearly independent, in order that $N_{1}^{\prime}$ can recover the vectors $\textbf{a}_{1}^{1},\textbf{a}_{1}^{2},\cdots,\textbf{a}_{1}^{\alpha}$. Conversely, suppose that they are linearly dependent. Without loss of generality, assume that  $\sum^\alpha_{i=1}d^{(1)}_{\gamma,i}\textbf{a}_{1}^{i}$ is a linear combination of the others. By the equations in (5), we easily deduce that $\textbf{a}_{k+1}^{\gamma}$ can be expressed linearly by the $(M-1)$ vectors $\textbf{a}_{2}^{1},\cdots,\textbf{a}_{2}^{\alpha}$, $\cdots,\textbf{a}_{k}^{1},\cdots,\textbf{a}_{k}^{\alpha},\textbf{a}_{k+1}^{1}$, $\cdots,\textbf{a}_{k+1}^{\gamma-1}$, which conflicts with the property of linear independence of these $M$ node vectors. Thus, $N_1$ can be repaired successfully, and the total repair bandwidth is $(k-1)\alpha+\gamma=M$. This completes the proof.
\end{IEEEproof}

\begin{eg}
Review that Example \ref{EX_2} provides an ORDSS of the unidirectional ring network with parameters $(n=4,\alpha=2,M=5)$. By Lemma \ref{lemma-iff}, this ORDSS satisfies the two conditions below: (i) all the four node vectors in arbitrary two adjacent storage nodes are linearly independent; (ii) arbitrary three adjacent storage nodes contain five linearly independent node vectors. In the following, using the repair method presented in the proof of Theorem \ref{thm-ORDSS-repair}, we propose an optimal repair process of the storage node $N_1$ with repair bandwidth 5. First, from storage nodes $N_1,N_2,N_3$, we choose five node symbols $x_1,x_1+6x_2+10x_3+5x_4+x_5,x_2,5x_1+9x_2+x_3+4x_4+4x_5,x_3$, whose corresponding node vectors are linearly independent. Then we represent them by new symbols $y_1,y_2,y_3,y_4,y_5$, that is,
\begin{equation*}\left\{
\begin{aligned}
y_1&\triangleq x_1, \\
y_2&\triangleq x_1+6x_2+10x_3+5x_4+x_5, \\
y_3&\triangleq x_2,\\
y_4&\triangleq 5x_1+9x_2+x_3+4x_4+4x_5,\\
y_5&\triangleq x_3.
\end{aligned}\right.
\end{equation*}
The remaining node symbols can be expressed linearly by the five new symbols as follows:
\begin{equation*}\left\{
\begin{aligned}
4x_1+7x_2+5x_3+2x_4+5x_5&=y_1+2y_2+2y_3+9y_4+9y_5, \\
x_4&=9y_1+3y_2+8y_3+2y_4+y_5,\\
x_5&=9y_1+8y_2+9y_3+y_4+7y_5.
\end{aligned}\right.
\end{equation*}
Fig. \ref{fig_ex2_rep} describes the storage scheme by using the new symbols, and Fig. \ref{fig_ex2_rep2} shows an optimal repair process of $N_1$.
\begin{figure}[!ht]
\centering
\includegraphics[width=10cm,height=4.8cm]{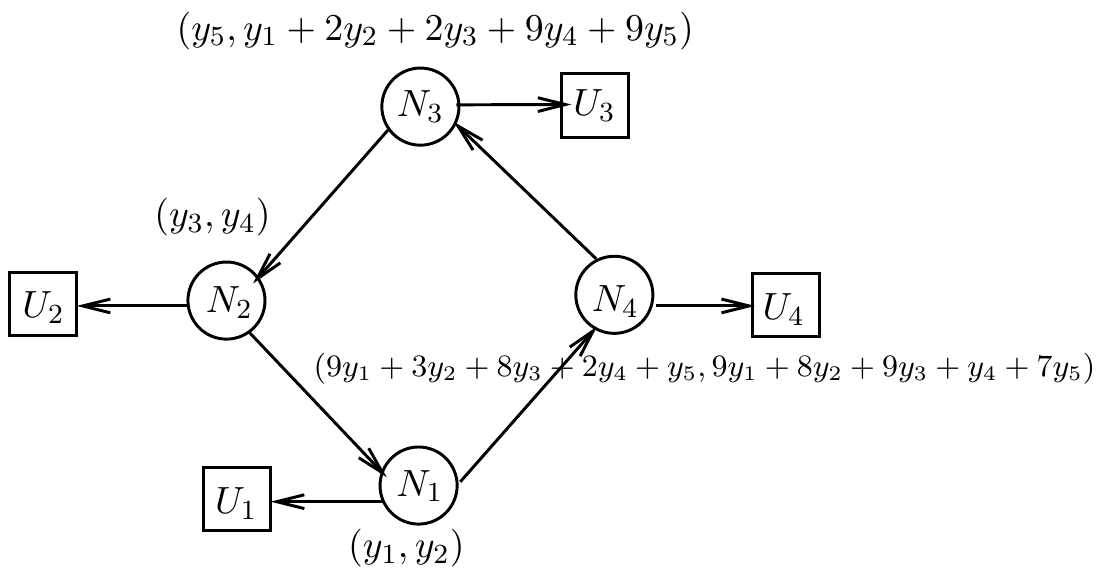}
\caption{The transformed storage scheme with the new symbols $y_1,y_2,y_3,y_4,y_5$.}
\label{fig_ex2_rep}
\end{figure}

\begin{figure}[!ht]
\centering
\includegraphics[width=14cm,height=5.1cm]{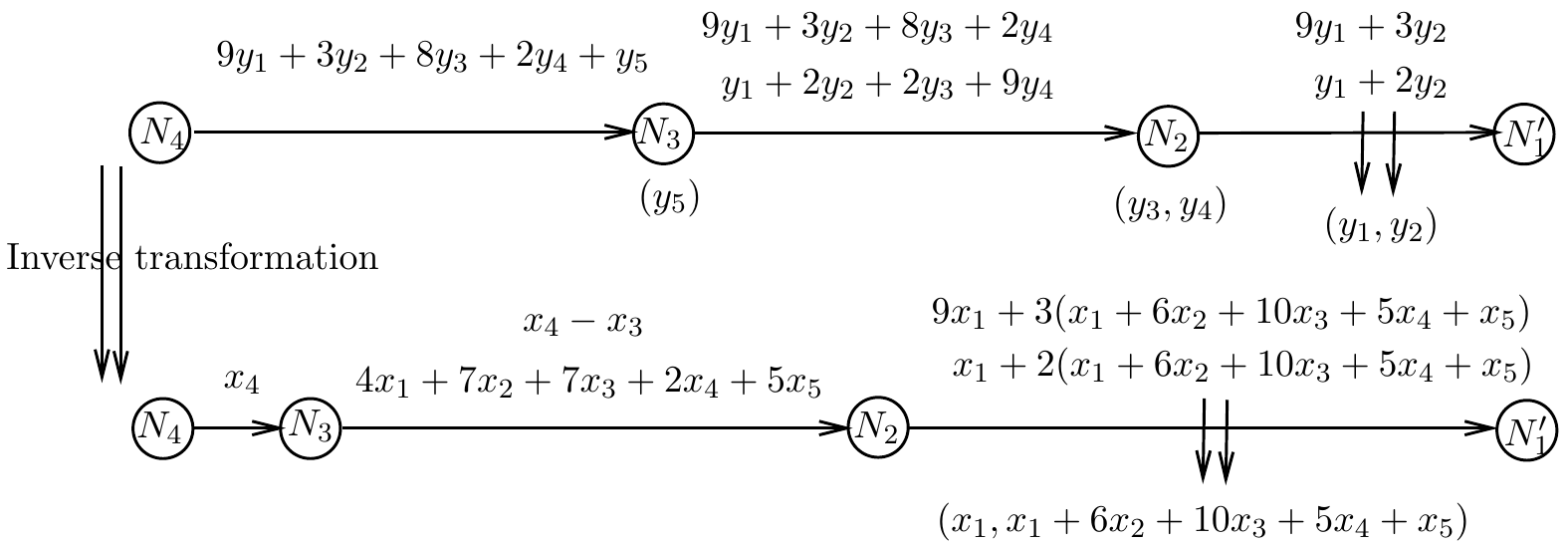}
\caption{The optimal repair process of the storage node $N_1$.}
\label{fig_ex2_rep2}
\end{figure}

As depicted in Fig. \ref{fig_ex2_rep2}, $N_3$ eliminates $y_5$ from the received symbol $9y_1+3y_2+8y_3+2y_4+y_5$, and $9y_5$ from its own stored symbol $y_1+2y_2+2y_3+9y_4+9y_5$. Then it transmits $9y_1+3y_2+8y_3+2y_4$ and $y_1+2y_2+2y_3+9y_4$ to $N_2$. With the stored symbols $y_3,y_4$ of $N_2$, $N_2$ eliminates
$8y_3+2y_4$ and $2y_3+9y_4$ from the two received symbols and transmits $9y_1+3y_2$ and $y_1+2y_2$ to $N_1^\prime$. Now $N_1^\prime$ can recover $y_1,y_2$ easily, which are exactly the symbols stored in $N_1$. The other storage nodes can be also repaired similarly with repair bandwidth $5$.
\end{eg}

Actually, our ring topology guarantees that, for ORDSSes, the minimum repair bandwidth for functional repair is no longer less than that for exact repair. Here, exact repair means that the stored data of the new substituted node must be the same as that in the failed node, while functional repair only needs to preserve the MDS property and the stored data of the new substituted node is not necessarily the same as that in the failed node, referring to \cite{A-Survey-2011} for details. In general, the minimum repair bandwidth for functional repair is less than that for exact repair. However, exact repair has many advantages than functional repair, such as lower computational complexity and stronger storage security applications in the face of eavesdroppers. By Theorem \ref{thm-ORDSS-repair}, it is easy to see that, for any ORDSS, every storage node can be repaired exactly when it fails. Moreover, the repair bandwidth achieves the lower bound $M$ in Theorem \ref{thm-repair}. Thus, it is sufficient to construct ORDSSes.


\section{Construction of ORDSSes}\label{cons}
In this section, we focus on an efficient construction of ORDSSes. Recall that the proof of Theorem \ref{thm-rec-bound} has provided a construction of ORDSSes, called MDS construction. However, this construction needs large finite field size, thereby has high computational complexity. In the following, we use the concept of Euclidean division to present another construction of ORDSSes, called ED construction, which has many advantages than MDS construction. Before the construction, we give some definitions and notation firstly.

\begin{defn}\label{def3}
For any two finite positive integers $M_0$ and $M_1$ with $M_1< M_0$, by Euclidean division, we have a series of equalities for some integer $k$:
\begin{equation*}\left\{
\begin{array}{ll}
M_0=P_1M_1+M_2, & 0< M_2<M_1,\\
M_1=P_2M_2+M_3, & 0< M_3<M_2,\\
\cdots & \cdots\\
M_{k-2}=P_{k-1}M_{k-1}+M_{k},& 0< M_{k}<M_{k-1},\\
M_{k-1}=P_{k}M_{k}.&
\end{array}\right.
\end{equation*}
According to the above equalities, we define an $M_1\times M_0$ matrix $G$ as follows:
\begin{equation*}
\left[
\begin{array}{ccc@{}|c@{}}
I_{M_1} & \cdots & I_{M_1} & \begin{array}{c}
I_{M_2}\\
\vdots\\
I_{M_2}\\\hline
\begin{array}{ccc|c}
I_{M_3} & \cdots & I_{M_3} & \begin{array}{c}
I_{M_4}\\
\vdots\\
I_{M_4}\\\hline
\cdots
\end{array}
\end{array}
\end{array}
\end{array}\right]_{M_1 \times M_0},
\end{equation*}
where $I_{M_i}$ represents the $M_i\times M_i$ identity matrix and the number of $I_{M_i}$ is $P_i$, $1 \leq i \leq k$. Such matrix $G$ is called ED-matrix.
\end{defn}

\begin{defn}\label{def-weakly MDS property}
For an $M \times N$ matrix, it is said to satisfy weak-column (or, weak-row) MDS property, if its arbitrary $M$ cyclic adjacent columns (or, arbitrary $N$ cyclic adjacent rows) are linearly independent when $M\leq N$ (or, $M>N$). Here, ``cyclic adjacent'' means that the last column (resp. row) of this matrix is regarded to be adjacent with the first column (resp. row).
\end{defn}

\begin{thm}\label{thm-MDS}
ED-matrices satisfy the weak-column MDS property.
\end{thm}

Please refer to Appendix \ref{app2} for the details of the proof.

In the following, we present the ED construction approach of ORDSSes.
For a unidirectional ring network with arbitrary parameters $(n,\alpha,M)$, we select an $M\times n\alpha$ ED-matrix $G$ as the generator matrix of its distributed storage scheme. Similarly, $X$ is the $M$-dimensional row vector of original data. Then assign the $n\alpha$ coordinates of the product $XG$ (equivalently, the $n\alpha$ column vectors of $G$) to $n$ storage nodes by the following approach: assign the first $\alpha$ coordinates to the storage node $N_1$, the second $\alpha$ coordinates to the storage node $N_2$, so far and so forth, the last $\alpha$ coordinates, i.e., the $n$th $\alpha$ coordinates, to the storage node $N_n$.

By Theorem \ref{thm-MDS}, we know that arbitrary $M$ cyclic adjacent columns of the above $M \times n\alpha$ generator matrix $G$ are linearly independent. Thus, the above assignment shows that the node vectors in arbitrary $k-1$ adjacent storage nodes are linearly independent and arbitrary $k$ adjacent storage nodes contain $M$ linearly independent node vectors, where $k=\lceil M/\alpha \rceil$. Together with Lemma \ref{lemma-iff}, this storage scheme is an ORDSS. Therefore, all users can recover the original data with the minimum reconstructing bandwidth $kM-\frac{(k-1)k\alpha}{2}$. In addition, Theorem \ref{thm-ORDSS-repair} shows that any storage node can be repaired with repair bandwidth $M$ if it fails.

Notice that ED construction can be used for arbitrary parameters $(n,\alpha,M)$. Particularly, when $n=M$, the generator matrix $G$ degrades to $[I_M,\cdots,I_M]$, which consists of $\alpha$ identity matrices of size $M\times M$. So what stored in each storage node are all uncoded symbols.  The reconstructing and repair processes do not need any coding operations in this case, which enormously reduces the computational cost.

\begin{eg}\label{EX_3}
For a unidirectional ring network with parameters $(n=4,\alpha=2,M=5)$, which is the same as that in Example \ref{EX_2}, let $X=[x_1, x_2, x_3, x_4, x_5]\in \mathbb{F}^5_2$ be the row vector of original data. We choose the ED-matrix $G$ of size $5 \times 8$ as follows:

$$G=\left[\begin{array}{cccccccc}1&0&0&0&0&1&0&0\\0&1&0&0&0&0&1&0\\
0&0&1&0&0&0&0&1\\0&0&0&1&0&1&0&1\\0&0&0&0&1&0&1&1\end{array}\right].$$

Subsequently, we calculate $$XG =[x_1, x_2, x_3, x_4, x_5, x_1+x_4, x_2+x_5, x_3+x_4+x_5].$$
Then, we assign $x_1,x_2$ to the storage node $N_1$, $x_3,x_4$ to the storage node $N_2$, $x_5, x_1+x_4$ to the storage node $N_3$ and $x_2+x_5$, $x_3+x_4+x_5$ to the storage node $N_4$. Clearly, every user can reconstruct the original data $X$ with reconstructing bandwidth 9. If any storage node fails, it can be repaired with repair bandwidth 5. Fig. \ref{ED-construction} depicts the optimal repair process in detail for this storage scheme.
For instance, if the storage node $N_2$ fails, $N_1$ transmits $x_1$ to $N_4$, then $N_4$ transmits $x_1,x_3+x_4+x_5$ to $N_3$, $N_3$ can recover $x_3,x_4$ and transmits them to the new substituted node $N_{2}^\prime$. So $N_2$ is repaired exactly with repair bandwidth 5.
\begin{figure}[!ht]
\centering
\includegraphics[width=10cm,height=3.7cm]{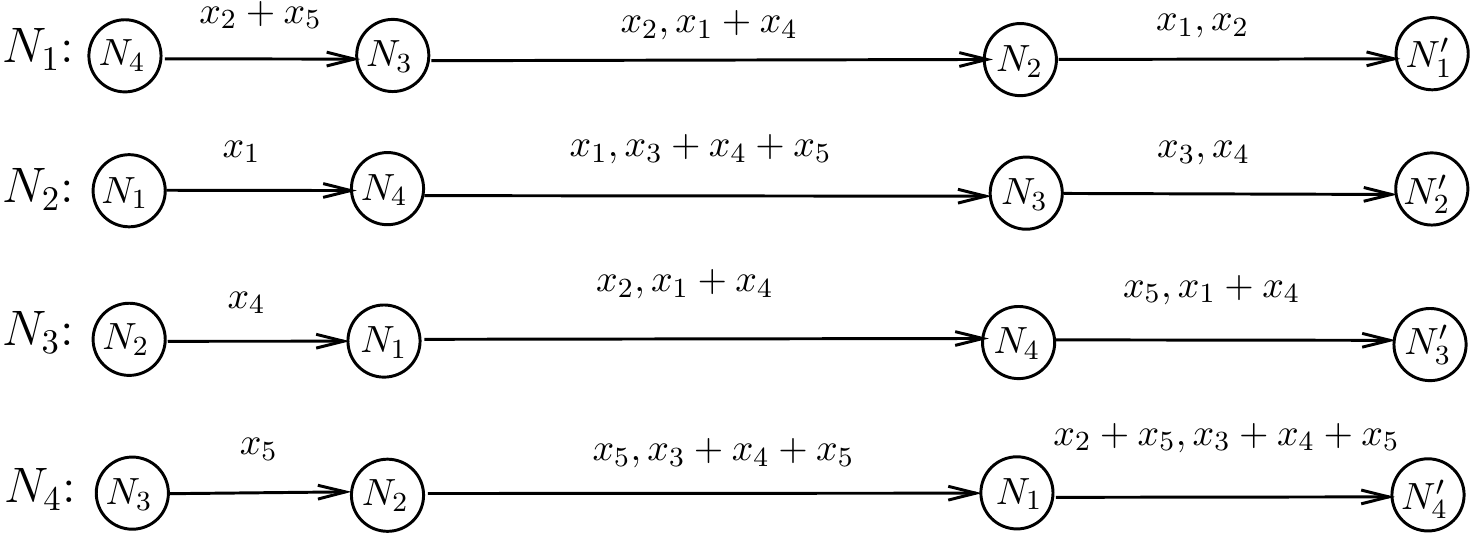}
\caption{The repair process for this storage scheme. }
\label{ED-construction}
\end{figure}
\end{eg}

Next, we compare the two constructions. MDS construction uses a generator matrix $G$ of an $[n\alpha, M]$ MDS code, whose arbitrary $M$ columns are linearly independent. This property, called MDS property, is too strong for constructing ORDSSes. The authors in \cite{M-S} have shown that the size of finite field $\mathbb{F}_q$ is not less than $n-k+1$ for the existence of an $[n,k,n-k+1]$ MDS code with $k\geq 2$. For MDS construction, when the parameters $(n,\alpha,M)$ of a unidirectional ring network take large values, the finite field size required will become too large for practical applications, while ED construction always uses the smallest finite field $\mathbb{F}_2$. This is because that the weakly MDS property is sufficient for constructing ORDSSes. For example, when $n=500,\alpha=10,M=1000$, the field size for a $[5000,1000]$ MDS construction is at least $5000-1000+1=4001$, which is much larger than the field size 2 for ED construction. It is well-known that the cost of arithmetic in a small field is smaller than that in a bigger one. Thus, the smaller field size will reduce the computational complexity of the storage scheme and save much time evidently. Therefore, ED construction is much better than MDS construction.

\section{Conclusion}\label{conc}
In this paper, we discuss distributed storage problems over unidirectional ring networks with parameters $(n,\alpha,M)$ and propose two tight lower bounds on the reconstructing and repair bandwidths. We define optimal reconstructing distributed storage schemes (ORDSSes). Particularly, we present two constructions for ORDSSes, called MDS construction and ED construction, respectively. Both of them can be used for arbitrary parameters $(n,\alpha,M)$, and ED construction is superior to MDS construction in terms of finite field size, computational complexity, etc. In practical applications, the networks of bidirectional ring topology, in which adjacent nodes can exchange data each other, are more useful. The same research problems in that case are also meaningful and still keep open.


\appendices
\section{Proof of lemma \ref{lemma-iff}}\label{app1}
\begin{IEEEproof}
Here, we still let $\textbf{a}_{i}^{1},\textbf{a}_{i}^{2},\cdots,\textbf{a}_{i}^{\alpha}$ be the $\alpha$ linearly independent node vectors of storage node $N_i$, $1\leq i\leq n$. First, we prove that if the two conditions are satisfied, then all users can recover the original data with the same reconstructing bandwidth $kM-\frac{(k-1)k\alpha}{2}$. Due to the symmetry of the network, it suffices to discuss the reconstructing bandwidth for the user $U_1$ connecting $N_1$ to download data. According to the two conditions, we know that all the $(k-1)\alpha$ node vectors in the storage nodes $N_1,N_2,\cdots,N_{k-1}$ are linearly independent, and the storage nodes $N_1,N_2,\cdots,N_{k}$ contain $M$ linearly independent node vectors. Without loss of generality, assume that all the $(k-1)\alpha$ node vectors in $N_{1},N_2,\cdots,N_{k-1}$ and the $\gamma$ node vectors $\textbf{a}_{k}^{1},\textbf{a}_{k}^{2},\cdots,\textbf{a}_{k}^{\gamma}$ in $N_{k}$ are $M$ linearly independent node vectors, where $\gamma=M-(k-1)\alpha$. In order to enable $U_1$ to reconstruct the original data, $N_{k}$ transmits $\textbf{a}_{k}^{1},\textbf{a}_{k}^{2},\cdots,\textbf{a}_{k}^{\gamma}$ to $N_{k-1}$, then $N_{k-1}$ sends the received $\gamma$ vectors and its own $\alpha$ node vectors to $N_{k-2}$. Continue this process until the storage node $N_1$. $N_1$ can receive $(k-2)\alpha+\gamma$ node vectors. Together with its own $\alpha$ node vectors, $N_1$ obtains total $M$ linearly independent node vectors. So it can recover the original data $X$ and output to the user $U_1$. Thus, the total reconstructing bandwidth can be calculated as follows:$$\gamma+[\gamma+\alpha]+\cdots+[\gamma+(k-2)\alpha]+M=(k-1)[M-(k-1)\alpha]+\frac{(k-2)(k-1)\alpha}{2}+M=kM-\frac{(k-1)k\alpha}{2}.$$

In the following, we indicate that if all users can recover the original data with the same reconstructing bandwidth $kM-\frac{(k-1)k\alpha}{2}$, then the two conditions should be satisfied. For ({\rom 1}), assume the contrary that all $(k-1)\alpha$ node vectors of some adjacent $k-1$ storage nodes are linearly dependent. Without loss of generality, suppose that $N_1,N_2,\cdots,N_{k-1}$ are these adjacent $k-1$ storage nodes and the number of the maximum linearly independent node vectors among them is $\widetilde{\alpha}$ with $\widetilde{\alpha}<(k-1)\alpha$. Thus, in order to reconstruct the original data, the user $U_1$ has to get another $M-\widetilde{\alpha}$ linearly independent vectors from other storage nodes $N_k,N_{k+1},\cdots,N_n$. By simple deduction, we can obtain that the reconstructing bandwidth for the user $U_1$ is not less than $kM-\frac{(k-1)k\alpha}{2}+((k-1)\alpha-\widetilde{\alpha})$, which makes a contradiction. For ({\rom 2}), suppose that the number of total linearly independent node vectors for some adjacent $k$ nodes is less than $M$. Without loss of generality, assume that $N_1,N_2,\cdots,N_k$ are these adjacent $k$ storage nodes and the number of total linearly independent node vectors is $\widetilde{M}$ with $\widetilde{M}<M$. Then the other $M-\widetilde{M}$ linearly independent vectors must be from other storage nodes $N_{k+1},N_{k+2},\cdots,N_n$. We cam easily deduce that the reconstructing bandwidth is not less than $kM-\frac{(k-1)k\alpha}{2}+(M-\widetilde{M})$, which contradicts to the fact that the reconstructing bandwidth is equal to $kM-\frac{(k-1)k\alpha}{2}$.
This completes the proof of the lemma.
\end{IEEEproof}


\section{Proof of Theorem \ref{thm-MDS}}\label{app2}

In the following, we will prove Theorem \ref{thm-MDS}, that is, ED-matrices satisfy the weak-column MDS property. Review that each ED-matrix has the following form:
\begin{equation*}
G=\left[
\begin{array}{ccc@{}|c@{}}
I_{M_1} & \cdots & I_{M_1} & \begin{array}{c}
I_{M_2}\\
\vdots\\
I_{M_2}\\\hline
\begin{array}{ccc|c}
I_{M_3} & \cdots & I_{M_3} & \begin{array}{c}
I_{M_4}\\
\vdots\\
I_{M_4}\\\hline
\cdots
\end{array}
\end{array}
\end{array}
\end{array}\right]_{M_1 \times M_0},
\end{equation*}
where $M_0> M_1>M_2>\cdots$.

Before the proof, we first start with a proposition.

\begin{prop}\label{weakly-MDS-prop1}
For the ED-matrix $G=[I_{M_1} \cdots I_{M_1}\mid G_1]$ as defined above, where $G_1$ is an $M_1 \times M_2$ matrix as follows:
\begin{equation*}
G_1=\left[
\begin{array}{c@{}}
 \begin{array}{c}
I_{M_2}\\
\vdots\\
I_{M_2}\\\hline
\begin{array}{ccc|c}
I_{M_3} & \cdots & I_{M_3} & \begin{array}{c}
I_{M_4}\\
\vdots\\
I_{M_4}\\\hline
\begin{array}{ccc|c}
I_{M_5} & \cdots & I_{M_5} & \vdots
\end{array}
\end{array}
\end{array}
\end{array}
\end{array}\right]_{M_1 \times M_2},
\end{equation*}
if $G_1$ satisfies the weak-row MDS property, then this ED-matrix $G$ satisfies the weak-column MDS property.
\end{prop}

In order to keep the continuity, the proof of Proposition \ref{weakly-MDS-prop1} is deferred to the end of this appendix.

\begin{IEEEproof}[Proof of Theorem \ref{thm-MDS}]
In order to show the weak-column MDS property of $G$, it is just needed to prove the weak-row MDS property of $G_1$ by Proposition \ref{weakly-MDS-prop1}. This is equivalent to prove the weak-column MDS property of $G_1^\top$, the transposition of $G_1$. Notice that $G_1^\top$ is also an ED-matrix and has the form
$G_1^\top=[I_{M_2} \cdots I_{M_2}\mid G_2]$, where
\begin{equation*}
G_2=\left[
\begin{array}{c@{}}
 \begin{array}{c}
I_{M_3}\\
\vdots\\
I_{M_3}\\\hline
\begin{array}{ccc|c}
I_{M_4} & \cdots & I_{M_4} & \begin{array}{c}
I_{M_5}\\
\vdots\\
I_{M_5}\\\hline
\begin{array}{ccc|c}
I_{M_6} & \cdots & I_{M_6} & \vdots
\end{array}
\end{array}
\end{array}
\end{array}
\end{array}\right]_{M_2 \times M_3}.
\end{equation*}
Hence, again by Proposition \ref{weakly-MDS-prop1}, it suffices to prove the weak-row MDS property of $G_2$, which is equivalent to prove the weak-column MDS property of $G_2^\top$. Similarly, note that $G_2^\top$ is also an ED-matrix with the form $G_2^\top=[I_{M_3} \cdots I_{M_3}\mid G_3]$, where
\begin{equation*}
G_3=\left[
\begin{array}{c@{}}
 \begin{array}{c}
I_{M_4}\\
\vdots\\
I_{M_4}\\\hline
\begin{array}{ccc|c}
I_{M_5} & \cdots & I_{M_5} & \begin{array}{c}
I_{M_6}\\
\vdots\\
I_{M_6}\\\hline
\begin{array}{ccc|c}
I_{M_7} & \cdots & I_{M_7} & \vdots
\end{array}
\end{array}
\end{array}
\end{array}
\end{array}\right]_{M_3 \times M_4}.
\end{equation*}
Thus, according to Proposition \ref{weakly-MDS-prop1}, in order to show the weak-column MDS property of $G_2^\top$, it is sufficient to indicate that $G_3$ satisfies the weak-row MDS property. This is equivalent to show the weak-column MDS property of $G_3^\top$.

Continuing this analysis process, since both $M_0$ and $M_1$ are finite, this process will stop at some step, for example, the $k$th step. It is not difficult to see that
$G^\top_{k-1}=\left[
\begin{array}{ccc|c}
I_{M_k} & \cdots & I_{M_k} & \begin{array}{c}
I_{M_{k+1}}\\
\vdots\\
I_{M_{k+1}}
\end{array}
\end{array}\right]$
and
$G_{k}=\left[
 \begin{array}{c}
I_{M_{k+1}}\\
\vdots\\
I_{M_{k+1}}
\end{array}\right].$ It is evident that $G_k^\top$ satisfies the weak-column MDS property, that is, arbitrary $M_{k+1}$ cyclic adjacent columns are linearly independent. Therefore, by the above iterative procedure, we conclude that the ED-matrix $G$ satisfies the weak-column MDS property. This accomplishes the proof.
\end{IEEEproof}

The remaining is the proof of Proposition \ref{weakly-MDS-prop1}.

\begin{IEEEproof}[Proof of Proposition \ref{weakly-MDS-prop1}]
For the $M_1\times M_0$ ED-matrix $G$ with $M_1 < M_0$, we prove its weak-column MDS property, that is, arbitrary $M_1$ cyclic adjacent columns are linearly independent. It suffices to verify that the matrices formed by arbitrary $M_1$ cyclic adjacent columns are invertible.
Review that $G=[I_{M_1} \cdots I_{M_1}\mid G_1]\triangleq [G_0\mid G_1]$, where $G_0=[I_{M_1} \cdots I_{M_1}]$. Let $T$ be the submatrix formed by $M_1$ cyclic adjacent columns of $G$, which must have one of the following four cases.

{\bf Case 1}: All the $M_1$ cyclic adjacent columns are from the submatrix $G_0$. Then the corresponding matrix $T$ has the form below:
\begin{equation*}
T=\left[
\begin{array}{c@{}|c@{}}
{\bf 0} & \begin{array}{c}
I_{M_1-a}\\
\end{array}  \\ \hline
\begin{array}{c}
I_a \\
\end{array} & {\bf 0}\\
\end{array}\right],
\end{equation*}
where $0\leq a< M_1$. Evidently, $T$ is invertible.

{\bf Case 2}: The considered submatrix $T$ is constituted by the last $b$ columns of $G_0$ and the first $(M_1-b)$ columns of $G_1$, where $ M_1-M_2\leq b\leq M_1-1$. Specifically, $T$ has the following form:
\begin{equation*}
T=\left[
\begin{array}{c@{}|c@{}}
{\bf 0} & \begin{array}{c}
I_{M_1-b} \\
\end{array}  \\ \hline
\begin{array}{c}
I_b\\
\end{array} & \ast\\
\end{array}\right],
\end{equation*}
where
$\left[
\begin{array}{c}
I_{M_1-b} \\ \hline
\ast
\end{array}\right]$
is the submatrix consisting of the first $(M_1-b)$ columns of $G_1$. Then, we make row operations on the above matrix $T$ to obtain $T^{\prime}$ below:
\begin{equation*}
T^{\prime}=\left[
\begin{array}{c@{}|c@{}}
{\bf 0} & \begin{array}{c}
I_{M_1-b} \\
\end{array}  \\ \hline
\begin{array}{c}
I_b\\
\end{array} & {\bf 0}\\
\end{array}\right].
\end{equation*}
It is evident that $T^{\prime}$ is invertible. Thus, $T$ is also invertible.

{\bf Case 3}: The considered cyclic adjacent submatrix $T$ is constituted by the last $c$ columns of $G_0$, all the $M_2$ columns of $G_1$ and the first $(M_1-M_2-c)$ columns of $G_0$, where $0\leq c\leq M_1-M_2-1$. Thus, this matrix $T$ has the form as follows:
\begin{equation*}
T=\left[
\begin{array}{c|c|c}
{\bf 0} & G_{11} & I_{M_1-M_2-c} \\ \hline
{\bf 0} & G_{12} & {\bf 0}\\ \hline
I_c & G_{13} & {\bf 0}
\end{array}\right],
\end{equation*}
where
$G_1=\left[
\begin{array}{c}
G_{11}\\ \hline
G_{12}\\ \hline
G_{13}
\end{array}\right].$ Since the matrix $G_1$  satisfies the weak-row MDS property, arbitrary $M_2$ cyclic adjacent rows of $G_{1}$ are linearly independent, which implies that the $M_2 \times M_2$ submatrix $G_{12}$ of $G_1$ is invertible. Therefore, it is not difficult to see that $T$ is also invertible.

{\bf Case 4}: The considered submatrix $T$ is constituted by the last $d$ columns of $G_1$ and the first $(M_1-d)$ columns of $G_0$, where $1 \leq d \leq M_2-1$. Thus, this matrix $T$ has the following form:
\begin{equation*}
T=\left[
\begin{array}{c|c|c}
{\bf 0} & I_{M_2-d} & {\bf 0}  \\ \hline
G^1_{d} & {\bf 0} & I_{M_1-M_2} \\ \hline
G^2_{d} & {\bf 0} & {\bf 0}
\end{array}\right],
\end{equation*}
where
$\left[
\begin{array}{c}
{\bf 0}\\ \hline
G^1_{d}\\ \hline
G^2_{d}
\end{array}\right]$ is the submatrix consisting of the last $d$ columns of $G_1$.
We exchange
$\left[
\begin{array}{c}
{\bf 0}\\ \hline
G^1_{d}\\ \hline
G^2_{d}
\end{array}\right]$
and
$\left[
\begin{array}{c}
I_{M_2-d}\\ \hline
{\bf 0}\\ \hline
{\bf 0}
\end{array}\right]$
 in the above matrix $T$ to obtain the matrix $T^{\prime}$ as follows:
\begin{equation*}
T^{\prime}=\left[
\begin{array}{c|c|c}
I_{M_2-d} & {\bf 0} & {\bf 0}  \\ \hline
{\bf 0} & G^1_{d} & I_{M_1-M_2} \\ \hline
{\bf 0} & G^2_{d} & {\bf 0}
\end{array}\right].
\end{equation*}
Then, make the proper row operations on the submatrix
$\left[
\begin{array}{c}
I_{M_2-d}\\ \hline
{\bf 0}\\ \hline
{\bf 0}
\end{array}\right]$ to establish a new matrix
$\left[
\begin{array}{c}
I_{M_2-d}\\ \hline
G^1_{M_2-d}\\ \hline
G^2_{M_2-d}
\end{array}\right]$, that is constituted by the first $(M_2-d)$ columns of $G_1$. Note that the first $(M_2-d)$ rows of the remaining submatrix
$\left[
\begin{array}{c|c}
 {\bf 0} & {\bf 0}  \\ \hline
 G^1_{d} & I_{M_1-M_2} \\ \hline
 G^2_{d} & {\bf 0}
\end{array}\right]$
are all zero row vectors. So under the above row operations, it keeps unchanged. Thus, making the same row operations on $T^{\prime}$, we can obtain the new matrix
\begin{equation*}
T^{\prime\prime}=
\left[
\begin{array}{c|c|c}
I_{M_2-d} & {\bf 0} & {\bf 0}  \\ \hline
G^1_{M_2-d} & G^1_{d} & I_{M_1-M_2} \\ \hline
G^2_{M_2-d} & G^2_{d} & {\bf 0}
\end{array}\right]=
\left[
\begin{array}{c@{}}
\begin{array}{c|c}
 G_1 & \begin{array}{c}
{\bf 0}\\\hline
I_{M_1-M_2}\\ \hline
{\bf 0}
\end{array}
\end{array}
\end{array}\right].
\end{equation*}
Furthermore, we rewrite $G_1$ as
$\left[
\begin{array}{c}
G_{11}\\ \hline
G_{12}\\ \hline
G_{13}
\end{array}\right]$, where $G_{11}$ of size $(M_2-d)\times M_2$, $G_{12}$ of size $M_2\times M_2$ and $G_{13}$ of size $d\times M_2$. So $T^{\prime\prime}$ can be rewritten as the following form:
\begin{equation*}
T^{\prime\prime}=\left[
\begin{array}{c|c}
G_{11} & {\bf 0}  \\ \hline
G_{12} &  I_{M_1-M_2} \\ \hline
G_{13} & {\bf 0}
\end{array}\right].
\end{equation*}
Since $G_1$ has the weak-row MDS property, its arbitrary $M_2$ cyclic adjacent rows are linearly independent. Particularly, the last $d$ rows and the first $(M_2-d)$ rows of $G_1$ are $M_2$ cyclic adjacent rows, which deduces that they are linearly independent, that is, the rows of $G_{11}$ and $G_{13}$ are linearly independent. Therefore, it is not difficult to see that $T^{\prime\prime}$ is invertible, which means that $T$ is also invertible.
Combining the above four cases, we accomplish the proof.
\end{IEEEproof}


\end{document}